\documentclass[a4paper,fleqn,usenatbib]{mnras}

\usepackage{newtxtext,newtxmath}

\usepackage[T1]{fontenc}
\usepackage{ae,aecompl}

\usepackage{graphicx}	
\usepackage{amsmath}	
\usepackage{amssymb}	
\usepackage{verbatim}
\usepackage{footmisc}
\usepackage{xcolor}
\usepackage{tablefootnote}

\title[Stellar Mass-Size Relation at $z~\mathtt{\sim}~1$ with HST]{HST/WFC3 grism observations of $z~\mathtt{\sim}~1$ clusters: The cluster vs. field stellar mass-size relation and evidence for size growth of quiescent galaxies from minor mergers}

\author[Matharu et al.]{
J. Matharu,$^{1\thanks{E-mail: j.matharu@ast.cam.ac.uk}}$
A. Muzzin,$^{2}$
G.B. Brammer,$^{3}$
R.F.J. van der Burg,$^{4}$
M.W. Auger,$^{1}$
\newauthor
P.C. Hewett,$^{1}$
A. van der Wel,$^{5,6}$
P. van Dokkum,$^{7}$
M. Balogh,$^{8}$
\hyperlink{https://orcid.org/0000-0001-6251-3125}{J.C.C. Chan},$^{9}$
\newauthor
\hyperlink{https://orcid.org/0000-0003-3921-2177}{R. Demarco},$^{10}$
D. Marchesini,$^{11}$
E.J. Nelson,$^{12}$
A. Noble,$^{13}$
\hyperlink{https://orcid.org/0000-0002-6572-7089}{G. Wilson}$^{9}$
and H.K.C. Yee$^{14}$
\\
$^{1}$Institute of Astronomy, University of Cambridge, Madingley Road, Cambridge, CB3 0HA, UK\\
$^{2}$York University, 4700 Keele Street, Toronto, ON, M3J 1P3, Canada\\
$^{3}$Cosmic Dawn Center, Niels Bohr Institute, University of Copenhagen, Juliane Maries Vej 30, DK-2100 Copenhagen \O, Denmark\\
$^{4}$European Southern Observatory, 85748, Garching bei M{\"u}nchen, Germany\\
$^{5}$Max-Planck Institut f{\"u}r Astronomie, K{\"o}nigstuhl 17, D-69117 Heidelberg, Germany\\
$^{6}$Sterrenkundig Observatorium, Universiteit Gent, Krijgslaan 281 S9, B-9000 Gent, Belgium\\
$^{7}$Astronomy Department, Yale University, 52 Hillhouse Ave, New Haven, CT 06511, USA\\
$^{8}$Department of Physics and Astronomy, University of Waterloo, Waterloo, ON, N2L 3G1, Canada\\
$^{9}$Department of Physics and Astronomy, University of California Riverside, 900 University Avenue, Riverside, CA 92521, USA\\
$^{10}$Departamento de Astronom\'ia, Facultad de Ciencias F\'isicas y Matem\'aticas,
Universidad de Concepci\'on, Concepci\'on, Chile\\
$^{11}$Department of Physics \& Astronomy, Tufts University, 574 Boston Avenue Suites 304, Medford, MA 02155, USA\\
$^{12}$Harvard-Smithsonian Center for Astrophysics, 60 Garden Street, Cambridge, MA 02138, USA\\
$^{13}$MIT Kavli Institute for Astrophysics and Space Research, 70 Vassar St, Cambridge, MA 02109, USA\\
$^{14}$Department of Astronomy and Astrophysics, University of Toronto, 50 St. George Street, Toronto, ON, M5S 3H4, Canada\\
}

\date{Accepted XXX. Received YYY; in original form ZZZ}

\pubyear{2018}

\begin{document}
\label{firstpage}
\pagerange{\pageref{firstpage}--\pageref{lastpage}}
\maketitle

\begin{abstract}
Minor mergers are thought to be responsible for the size growth of quiescent field galaxies with decreasing redshift. We test this hypothesis using the cluster environment as a laboratory. Satellite galaxies in clusters move at high velocities, making mergers between them rare. The stellar mass-size relation in ten clusters and in the field is measured and compared at $z~\mathtt{\sim}~1$. Our cluster sample contains 344 spectroscopically-confirmed cluster members with Gemini/GMOS and 182 confirmed with HST WFC3 G141 grism spectroscopy. On average, quiescent and star-forming cluster galaxies are smaller than their field counterparts by ($0.08\pm0.04$)~dex and ($0.07\pm0.01$)~dex respectively. These size offsets are consistent with the average sizes of quiescent and star-forming field galaxies between \mbox{$1.2\leqslant z\leqslant1.5$}, implying the cluster environment has inhibited size growth between this period and $z~\mathtt{\sim}~1$. The negligible differences measured between the $z~\mathtt{\sim}~0$ field and cluster quiescent mass-size relations in other works imply that the average size of quiescent cluster galaxies must rise with decreasing redshift. Using a toy model, we show that the disappearance of the compact cluster galaxies might be explained if, on average, $\mathtt{\sim}40\%$ of them merge with their brightest cluster galaxies (BCGs) and $\mathtt{\sim}60\%$ are tidally destroyed into the intra-cluster light (ICL) between \mbox{$0\leqslant z\leqslant1$}. This is in agreement with the observed stellar mass growth of BCGs between \mbox{$0\leqslant z\leqslant1$} and the observed ICL stellar mass fraction at $z~\mathtt{\sim}~0$. Our results support minor mergers as the cause for the size growth in quiescent field galaxies, with cluster-specific processes responsible for the similarity between the field and cluster quiescent mass-size relations at low redshift.
\end{abstract}




\section{Introduction}
\label{introduction}

Observations of massive galaxies across a range of redshifts have now established that as the Universe ages, massive galaxies grow in size more than they do in stellar mass (e.g \citealt{Huertas-Company2013,VanderWel2014}). This disproportionately larger growth in size has been the topic of many recent studies attempting to pin-down the physical process responsible for driving it. Thus far, the majority of these studies argue that minor mergers (mergers between two galaxies that have a mass ratio of $>$~10:1 or in some cases $>$~3:1) can provide a plausible explanation to these observations \citep{Bezanson2009,Hopkins2009b,Naab2009,Trujillo2011,Hilz2012,Oser2012,Bluck2012,Ferreras2014}. Despite these findings, there is debate on whether minor mergers can fully account for this size growth. It has been suggested that much of the size growth in the quiescent population of galaxies can also be explained by the appearance of newly quenched star-forming galaxies (which have larger sizes than quiescent galaxies) that formed at later epochs \citep{Carollo2013}. Major mergers (equal-mass mergers) can also increase the size of galaxies, but they lead to a proportionate increase in stellar mass. If major mergers were the dominant route for galaxy size growth, there would be more high mass galaxies than currently observed at low redshifts \citep{Bezanson2009,Lopez-Sanjuan2009}. Major mergers are therefore unlikely to be the main drivers of galaxy size growth. Feedback from active galactic nuclei (AGN) and stellar winds can also lead to the expansion of a galaxy \citep{Fan2008}. However, the amount of mass removed in this process would have to be fine-tuned to reach the levels of galaxy size growth observed \citep{Bezanson2009}. 

One aspect that can alter this picture of galaxy size growth over time is the influence of environment. Galaxies residing in high-density environments such as clusters have higher peculiar velocities than galaxies residing in the low density field environment. The high velocity dispersions associated with clusters make mergers between satellite galaxies a rare occurrence \citep{Merritt1985,Delahaye2017}. It is therefore possible to use the cluster environment as a laboratory to test the currently favoured suggestion that minor mergers dominate the size growth of massive galaxies. Since minor mergers are expected to increase galaxy size more than they do stellar mass, the most direct way to test this is to measure the stellar mass-size relations in both environments at fixed redshift and compare them to see if there is a significant offset in size. This simple approach does not take into account other sub-dominant cluster-specific processes that may also alter the sizes and size distribution of cluster galaxies such as accelerated quenching, galaxy harassment (nearby high-speed encounters between satellite galaxies in clusters e.g. \citealt{Moore1996,Moore1998}) and mergers with the Brightest Cluster Galaxy (BCG) that would remove galaxies from the sample. In this simple scenario, if the predictions of minor mergers driving galaxy size growth are true, cluster galaxies should find themselves inhibited from size growth and will therefore be significantly smaller than field galaxies at fixed redshift.

Previous works that have attempted to do this find a range of results, making physical interpretation difficult. At low redshifts ($z<0.2$), \cite{Weinmann2009}, \cite{Maltby2010}, \cite{Lorenzo2013} and \cite{Cappellari2013} find the stellar mass-size relation of early-type galaxies to differ by $\leqslant20\%$ with environment, explicitly stating their results are not significant. \cite{Huertas-Company2013,Huertas-Company2013a} find that this difference has to be $<30-40\%$, in-line with the errors on their size measurements. All these works state their measurements are consistent with no size difference being present for early-type galaxies with environment. Similarly, \cite{Poggianti2013} found that old compact early-type galaxies are approximately $\leqslant33\%$ smaller in clusters compared to the field. \cite{Cebrian2014} find a much smaller difference of early-type cluster galaxies being $\mathtt{\sim}3.5\%$ smaller than their field counterparts,  but at high statistical significance ($>4\sigma$). Late-type galaxies are more often found to be smaller in clusters. \cite{Weinmann2009}, \cite{Maltby2010} and \cite{Lorenzo2013} find them to be smaller in clusters by $\geqslant15\%$, with the latter two studies claiming statistically significant results. \cite{Cebrian2014} also find late-type galaxies to be smaller in clusters at high statistical significance, but by a much smaller amount ($\mathtt{\sim}7.5\%$). At intermediate to high redshifts ($z\geqslant0.2$), \cite{Cooper2012}, \cite{Papovich2012a}, \cite{Bassett2013}, \cite{Lani2013}, \cite{Delaye2014a}, \cite{Chan2018} and \cite{Andreon2018} find early-type galaxies to be larger in clusters by $\geqslant24\%$. \cite{Rettura2010}, \cite{Newman2014b}, \cite{Sweet2016} and \cite{Morishita2016} however find early-type galaxies to differ in size by $<20\%$ between environments, consistent with no size difference. On the other hand, \cite{Valentinuzzi2010a,Valentinuzzi2010b} find cluster galaxies to be smaller than field galaxies by $\mathtt{\sim}25\%$. A summary of cluster versus field mass-size relation measurements (since $2009$) for quiescent galaxies can be found in Appendix~\ref{previous}.

The biggest challenge that comparative studies of the stellar mass-size relation with environment face is the task of minimising systematics between measurements made from different samples. It is often the case that field and cluster datasets were taken with different instruments, using different filters and analysed using different techniques. It is imperative to compare galaxy sizes at the same rest-frame wavelength, since aspects such as colour gradients across galaxies can vary differently in different rest-frames, altering size measurements. High redshift studies face the added difficulty of obtaining statistically significant sample sizes with high enough signal-to-noise ratio and resolution to measure galaxy sizes reliably.

In this work, we attempt to test the hypothesis that minor mergers drive the majority of galaxy size growth by comparing the stellar mass-size relations in the cluster and field environments at $z~\mathtt{\sim}~1$. We aim to overcome shortcomings and challenges faced by previous studies in a number of ways. Firstly, we use the largest spectroscopically-confirmed sample of star-forming and quiescent cluster galaxies at $z~\mathtt{\sim}~1$ to date. Secondly, we purposefully conduct observations and data reduction in an analogous fashion to the 3D-HST/CANDELS survey, which forms our comparative field sample. Finally, we use data with high enough signal-to-noise ratio and resolution obtained using the {\it Wide Field Camera 3} (WFC3) on the {\it Hubble Space Telescope} (HST) to measure galaxy sizes reliably.

This paper is organised as follows. In Section~\ref{methodology} we describe our cluster and field samples, our size determination method and the calculation of our stellar masses. We then go on to explain how our final sample for this comparative study is selected. In Section~\ref{grism_accuracy}, we present our findings on the precision and accuracy of grism-derived redshifts in helping to boost our cluster sample size. We then present the cluster mass-size relation at $z~\mathtt{\sim}~1$ and investigate its offset with the field relation in Section~\ref{mass_size_sec}. In Section~\ref{morphology} we explore whether differences in morphological composition drive the differences in the stellar mass-size relation between the two environments. In Section~\ref{BCG_growth}, we investigate how the distinct location of a BCG at the bottom of the cluster potential well can help constrain the dominant physical process driving galaxy evolution in clusters. Finally, we summarise our findings in Section~\ref{conclusions}.

All magnitudes quoted are in the AB system and we assume a $\Lambda$CDM cosmology with $\Omega_{m}=0.307$, $\Omega_{\Lambda}=0.693$ and $H_{0}=67.7$~kms$^{-1}$~Mpc$^{-1}$ \citep{Planck2015}.

\section{Methodology}
\label{methodology}

\subsection{Data}
\subsubsection{Cluster sample}
\label{cluster_sample}

\begin{table*}
	\centering
	\caption{The ten GCLASS clusters used in this study. For full names of the clusters, we refer to \protect\cite{Muzzin2012}. $R_{200}$ is the radius at which the mean interior density is 200 times the critical density of the Universe. $M_{200}$ is the mass enclosed within this radius. $\sigma_{v}$ is the line-of-sight velocity dispersion (see \protect\citealt{Biviano2016}). Note: The numbers listed in the last three columns indicate total numbers {\it before} / {\it after} quality checks relevant to this study are applied.}
	\label{tab:example_table}
	\begin{tabular}{lccccccccc} 
		\hline
		Name & $z_{spec}$ & $M_{200}$ & $R_{200}$ & $\sigma_{v}$ & Spec-{\it z} Members & Grism-{\it z} & Total Members\\
                 &  & [$10^{14}M_{\odot}$] & [kpc] & [km s$^{-1}$] & in HST FOV & Members & in HST FOV\\
		\hline
		SpARCS-0034 & 0.867 &  $2.0\pm0.8$ & $888\pm110$ & $609_{-66}^{+75}$ & 37/33 & 17/16 & 54/49\\
		SpARCS-0035 & 1.335 &  $5\pm2$ & $977\pm154$ & $941_{-137}^{+159}$ & 22/22 & 26/18 & 48/40\\
		SpARCS-0036 & 0.869 &  $5\pm2$ & $1230\pm129$ & $911_{-90}^{+99}$ & 43/41 & 31/30 & 74/71\\
                SpARCS-0215 & 1.004 &  $3\pm1$ & $953\pm103$ & $758_{-77}^{+85}$ & 42/40 & 32/28 & 74/68\\
		SpARCS-1047 & 0.956 &  $3\pm1$ & $926\pm138$ & $680_{-86}^{+98}$ & 22/19 & 15/14 & 37/33\\
		SpARCS-1051 & 1.035 &  $1.2\pm0.5$ & $705\pm102$ & $530_{-65}^{+73}$ & 34/32 & 11/10 & 45/42\\
		SpARCS-1613 & 0.871 &  $13\pm3$ & $1663\pm130$ & $1232_{-93}^{+100}$ & 73/65 & 41/25 & 114/90\\
                SpARCS-1616 & 1.156 &  $3\pm1$ & $854\pm107$ & $701_{-73}^{+81}$ & 38/35 & 19/17 & 57/52\\
		SpARCS-1634 & 1.177 &  $4\pm2$ & $1008\pm131$ & $835_{-82}^{+91}$ & 38/34 & 15/15 & 53/49\\
                SpARCS-1638 & 1.196 &  $1.9\pm0.9$ & $769\pm117$ & $585_{-65}^{+73}$ & 26/23 & 13/9 & 39/32\\
		\hline
	\end{tabular}
\end{table*}

\begin{figure*}
	\includegraphics[width=\textwidth]{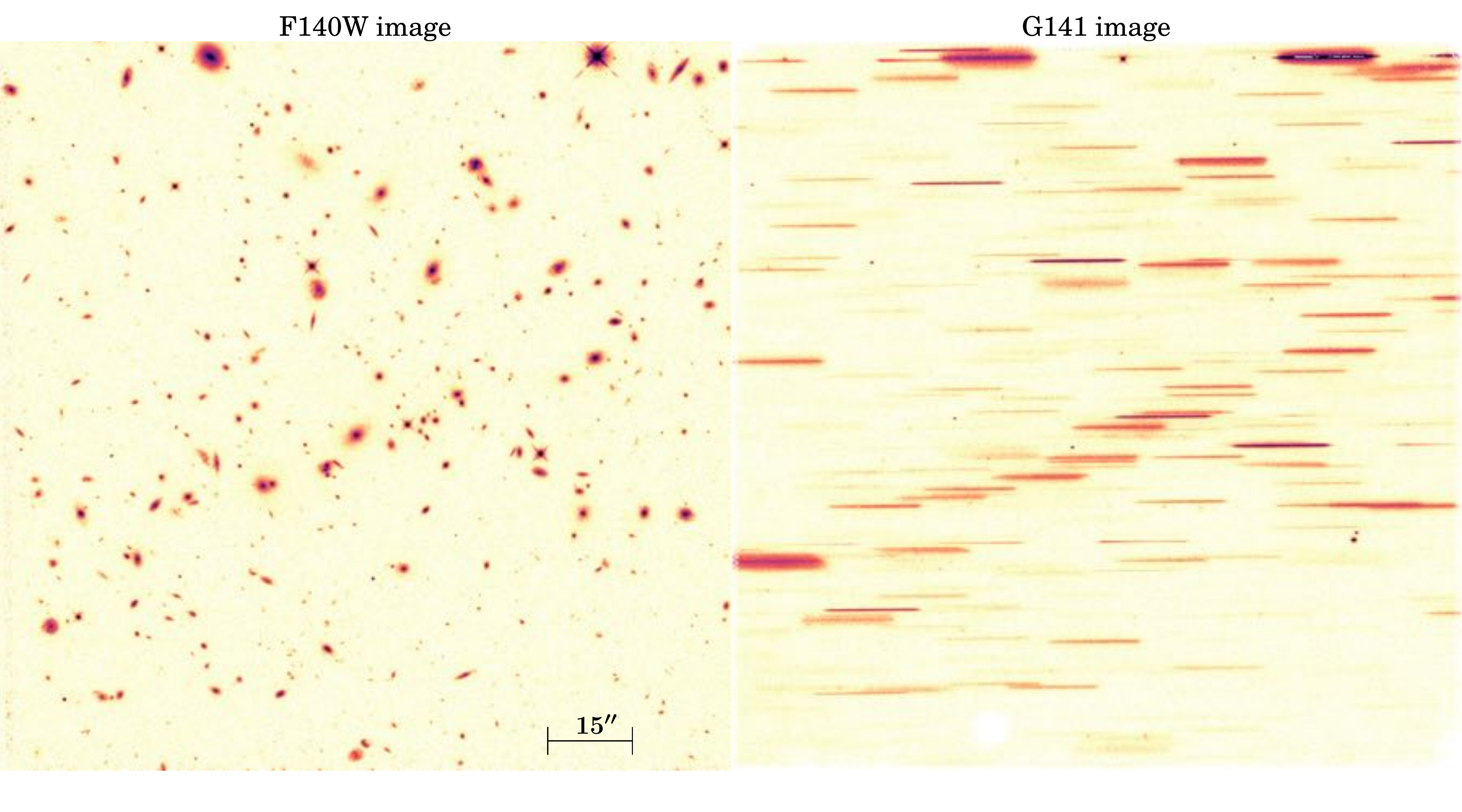}
    \caption{HST WFC3 data for SpARCS-1047 (see Table~\ref{tab:example_table}). Left: F140W image. Right: G141 grism spectra. Spatial scale of the F140W image is shown with a black bar in the bottom right-hand corner of the left panel. The colourmap is logarithmic. The grism provides a spatially resolved spectrum for every object in the field of view. For this particular cluster, $30\%$ of grism-selected cluster members were too contaminated to use (see Section~\ref{grism_contam} for contamination criteria).}
    \label{fig:grism_eg}
\end{figure*}

Our cluster sample consists of 10 massive clusters in the redshift range $0.86<z<1.34$ (see Table~\ref{tab:example_table}). These were selected from the 42 square degree {\it Spitzer} Adaptation of the Red-sequence Cluster Survey (SpARCS, see \citealt{Muzzin2009}; \citealt{Wilson2009a} and \citealt{Demarco2010}) for spectroscopic follow-up, as part of the Gemini Cluster Astrophysics Spectroscopic Survey (GCLASS, see \citealt{Muzzin2012} and \citealt{VanderBurg2013}). Clusters in the SpARCS survey were selected with the red-sequence method \citep{Gladders2000} using the $z^{\prime}$ - 3.6$\mu$m colour to sample the 4000{\AA} break (see \citealt{Muzzin2008}). The 10 clusters selected as part of GCLASS have extensive optical spectroscopy, obtained using the Gemini Multi-Object Spectrographs (GMOS) on both Gemini-South and -North. Out of 1282 galaxies that obtained a spectroscopic redshift, 457 were identified as cluster members. All 10 clusters also have 11-band photometry ({\it ugrizJK$_{s}$}, 3.6$\mu$m, 4.5$\mu$m, 5.8$\mu$m, 8.0$\mu$m), details of which can be found in Appendix A of \mbox{\cite{VanderBurg2013}}. An additional band of photometry in F140W is now also available as part of the HST data the study in this paper is based on.

\subsubsection{HST observations}
\label{hst_observations}
We obtained HST WFC3 F140W imaging and G141 grism follow-up in a Cycle 22 program (GO-13845; PI Muzzin), dedicated to obtaining spatially resolved H$\alpha$ maps of the star-forming cluster members. The F140W filter spans the wide {\it JH} wavelength range \mbox{($12003<~\lambda~/$~{\AA}~$<15843$)}. The G141 grism covers a wavelength range of \mbox{$10750<~\lambda~/$~{\AA}~$<17000$}, and therefore contains the H$\alpha$ emission line between  \mbox{$0.7<z<1.5$}. It provides a spatially resolved spectrum for every object in the field of view, with a resolving power of \mbox{$R~\mathtt{\sim}~130$}. 

Data for 9 of the 10 clusters are taken with a $1\times2$ or $2\times1$ mosaic of WFC3 pointings with a random orient. For the 10th cluster, SpARCS-1047 (see Table~\ref{tab:example_table}), data is taken with a single pointing at the centre of the cluster. All clusters are observed to a 2-orbit depth, with $\mathtt{\sim}90\%$ of this time spent on grism spectroscopy and $\mathtt{\sim}10\%$ on direct imaging with the F140W filter. The exposure time in F140W for 9 of the 10 clusters is $\mathtt{\sim}800$~seconds, with a 5$\sigma$ F140W limiting magnitude for galaxies of $\mathtt{\sim}26.6$. The exception to this is SpARCS-0035, which has varying depth, ranging from $\mathtt{\sim}800$ to $\mathtt{\sim}7460$ seconds\footnote{This cluster is being used for The Supernova Cosmology Project ``See Change" program, details of which can be found in \cite{Rubin2017}. As a result, this cluster's observations were deeper, with pointings of different orientations overlapping each other.\label{seechange}}. All mosaics are centered on the cluster core and cover $\mathtt{\sim}~8$~arcmin$^{2}$ for each cluster. This coincides with approximately a quarter of the GMOS observing area, which covered 35 arcmin$^2$ per cluster. For all of the GCLASS clusters, the HST imaging covers most if not all of the area within $R_{500}$. Four of the GCLASS clusters (SpARCS-0035, SpARCS-1616, SpARCS-1634 and SpARCS-1638; See Table~\ref{tab:example_table}) have imaging going out to $\mathtt{\sim}~R_{200}$. Nevertheless, the spectroscopic density is highest in the core, leading to a far greater overlap with spectroscopically confirmed cluster members. 392 of the 457 spectroscopically confirmed cluster members from GCLASS are in the HST fields-of-view (see Table~\ref{tab:example_table}). Figure~\ref{fig:grism_eg} shows the F140W and G141 images for one of the GCLASS clusters for illustration.

\subsubsection{Data reduction}
\label{data_reduction}
The F140W imaging consists of 20 SPARS50 readouts per cluster with a native pixel scale of $0.12^{\prime\prime}$ per orbit. These are reduced using the same data reduction process as for 3D-HST \citep{Brammer2012,Momcheva2016}, which forms our field sample (see Section~\ref{field_sample}). In summary, the standard calibrated data products from the HST archive are used with reduction pipelines to flag bad pixels, subtract bias structure, dark current and apply flat-field corrections. The $\mathtt{MultiDrizzle}$ software is then used to identify cosmic rays and hot pixels missed by these pipelines. The PyRAF routine $\mathtt{tweakshifts}$ is used to adjust the mosaics for dither offsets and the World Coordinate System (WCS) coordinates are subsequently refined to match the locations of objects on the mosaics. To subtract the sky background, a second-order polynomial fit is subtracted from each exposure after masking objects detected in the $\mathtt{MultiDrizzle}$ mosaics, and mapped back to the original frame using the PyRAF routine $\mathtt{blot}$. This entire process is explained in detail in \cite{Brammer2012}, and leads to undistorted, sky-subtracted F140W mosaics with a pixel scale of $0.06^{\prime\prime}$. The $\mathtt{MultiDrizzle}$ process also outputs weight maps for each mosaic. These provide values for the inverse variance at each pixel, making them integral to the noise estimation process when determining the sizes of cluster galaxies (see Section~\ref{noise_estimation}).

\subsubsection{Grism spectra contamination}
\label{grism_contam}
One of the major concerns regarding grism data for high-density fields such as clusters is the contamination rate. Our primary criterion for evaluating contamination is anything that inhibits the ability to obtain an accurately measured redshift from the grism spectrum. Therefore, a grism spectrum is considered unusable if one or more of the following criteria are true:
\begin{enumerate}
\item Less than half of the grism spectrum is available in a region where prominent emission/absorption lines are absent. 
\item The grism spectrum of a brighter close-by object overlaps with the grism spectrum of interest to such an extent that the data reduction process could not successfully remove the contamination.
\item The grism spectrum has a very low signal-to-noise ratio such that no prominent emission/absorption lines can be detected. This is often the case for galaxies with F140W magnitude~$>25$.
\end{enumerate}
We checked all the grism spectra by eye for galaxies without a spectroscopic redshift from GMOS, but with a grism redshift measurement placing them in the cluster. Approximately $35\%$ of grism-selected cluster members (see Section~\ref{grism_accuracy}) were too contaminated to use. $2\%$ of these contaminated grism spectra were unusable due to the galaxy being too close to the edge of the field-of-view, leaving less than half a grism spectrum available.

\subsubsection{Field sample}
\label{field_sample}
The HST observations for GCLASS are purposely conducted in an analagous fashion to the 3D-HST survey to allow for the most straightforward comparison between galaxies in field and cluster environments. The 3D-HST survey is a 248-orbit near-infrared spectroscopic treasury program, covering three quarters of the CANDELS treasury survey area with 2 orbits of WFC3 and G141 grism coverage \citep{Brammer2012}. It obtained rest-frame optical spectra for galaxies between \mbox{$0.7<z<3.5$} \citep{Brammer2012,Momcheva2016}. 3D-HST has obtained $\mathtt{\sim}100,000$ galaxy redshift measurements of which $\mathtt{\sim}10,000$ reside in a redshift range relevant to our work \citep{Momcheva2016}. In this study, we use the structural parameters for all galaxies measured by the 3D-HST team within the same redshift range as the GCLASS clusters \mbox{($0.86<z<1.34$)}, derived from the CANDELS \citep{Grogin2011,Koekemoer2011} F160W images for the COSMOS, UDS, GOODS-S, GOODS-N and AEGIS fields in \cite{VanderWel2012}.

Furthermore, we apply our size determination method for the GCLASS clusters to the CANDELS-COSMOS F160W mosaic\footnote{The CANDELS-COSMOS F160W mosaic (as well as the other CANDELS F160W images) is deeper than the GCLASS F140W mosaics. Therefore, there is a possibility that sizes measured from the CANDELS-COSMOS F160W mosaic are systematically larger due to a higher signal-to-noise ratio than the GCLASS F140W mosaics. We ran our size determination method on the 3D-HST COSMOS F140W mosaic which has the same depth as the GCLASS F140W mosaics. We compared our F160W and F140W sizes for the same set of galaxies in COSMOS, finding in fact a systematic towards larger sizes in the 3D-HST COSMOS F140W mosaic (see Appendix~\ref{filter_difference} for more details).\label{depth}}. This is done to ensure comparable sizes are achieved with those in \cite{VanderWel2012}, confirming the reliability of our method (see Appendix~\ref{size_agreement} for further discussion).

\subsection{Redshifts \& stellar masses}
\label{photometric_redshifts}
Part of obtaining a robust measurement of the stellar mass-size relation depends heavily upon reliably estimated stellar masses from fitting multi-wavelength Spectral Energy Distributions (SEDs) to the photometry. Once again, to minimise systematics between measurements from the cluster and field samples, we use the same method for the clusters as used for 3D-HST.

Photometric redshifts for all galaxies in the GCLASS mosaics are estimated using the EAZY code \citep{Brammer2008}. These are primarily used as a comparison to our spectroscopic and grism-derived redshifts (see Section~\ref{grism_accuracy}). They are not used for any quantitative measurements regarding the cluster mass-size relation in this study. The spectroscopic redshifts in this study were obtained with GMOS as part of the GCLASS survey (see Section~\ref{cluster_sample}). The grism redshifts are derived from fitting the SEDs with the grism spectra obtained with the G141 grism on the HST WFC3. The P$(z)$ obtained from the EAZY run for each galaxy is used as a prior to obtaining the grism redshift measurement.

Stellar masses are estimated using the FAST code \citep{Kriek2009}. Once the final selection of cluster members are made for this study using grism redshifts (see Section~\ref{sample_selection}), stellar masses for the cluster members are estimated with their redshifts fixed to the spectroscopic redshift of their respective cluster. We use the exponentially declining star formation history parameterisation, \mbox{SFR $\propto e^{-t/\tau}$}, where $\tau$ can range between 10 Myr and 10 Gyr and \mbox{$0<A_{V}<4$}. A \mbox{\cite{Chabrier2003}} Initial Mass Function (IMF) is assumed, as well as solar metallicity and the \cite{Calzetti1999} dust law.

Stellar masses for 3D-HST were estimated in exactly the same way as for GCLASS (with redshifts fixed to their spectroscopic redshifts or photometric redshifts when spectroscopic redshifts were not available), except that the allowed minimum value for $\tau$ was set to 40 Myr as opposed to 10 Myr \citep{Skelton2014a}.

The final stellar masses for both 3D-HST and GCLASS are corrected for the difference between the total F160W (or F140W) flux from the photometric catalogue and the total F160W (or F140W) flux as measured by GALFIT. This is done to ensure both stellar mass and size measurements are based on the same model for the galaxy light profile.

\subsection{Size determination}
\label{size_determination}
In this section we describe our methodology for measuring sizes and how we treat systematics at each stage. Our size determination method very closely follows the method used by \cite{VanderWel2012} for 3D-HST. Our size determination process is first tested on the CANDELS-COSMOS F160W mosaic\footref{depth} from the field sample. Our results are then compared to published results from \cite{VanderWel2012} in order to verify the reliability of our method (see Appendix~\ref{size_agreement}). Using the same size determination method for both cluster and field is plausible given our data. This is because there is negligible crowding of sources in the cluster mosaics compared to the field mosaics and the surface brightness of the intra-cluster light (ICL) is very low for the clusters.

\subsubsection{Source detection}
SExtractor v2.19.5 \citep{Bertin1996} is used to identify objects in the GCLASS F140W and CANDELS-COSMOS F160W mosaics. The SExtractor set-up for GCLASS is standard, details of which can be found in Appendix~\ref{GCLASS_sex}. For COSMOS we use the method described in Appendix A of \cite{Galametz2016}. In summary, this method runs SExtractor in a ``hot'' and ``cold'' mode separately, optimised for detecting faint and bright sources respectively. The two separate catalogues are then combined, ensuring no repeat detections. This method minimises the chances that a galaxy is split into mutliple objects as well as ensuring neighbouring galaxies are deblended adequately.

\subsubsection{Noise estimation}
\label{noise_estimation}
We construct noise maps for the GCLASS images and CANDELS-COSMOS mosaic in the same way noise maps were constructed in \cite{VanderWel2012}. These consist of a background noise estimate plus the Poisson noise from the sources themselves. The SExtractor segmentation maps are used in conjunction with the drizzled weight images from the data reduction process (see Section~\ref{data_reduction}) to select all the pixels that are not occupied by sources or detector defects. The root-mean-square value of the background pixels for each image is our estimate for the background noise per pixel in electrons~s$^{-1}$. The Poisson noise from the sources is calculated using the exposure time maps and science images. The units of the science image pixels are converted to electrons before the Poisson noise is calculated. We then convert this back to electrons~s$^{-1}$ and add this to our background noise map. The exception to this method of noise estimation is SpARCS-0035. Due to this cluster's non-trivial mosaic\footref{seechange} -- leading to heterogeneous noise properties -- the noise estimation is calculated using the weight image, where the noise, $\sigma = 1/\sqrt{weight}$ at each pixel in electrons~s$^{-1}$. This noise estimation is usually an underestimation, since it does not include the Poisson term from the sources. In practice, this only becomes a problem for the brightest sources which are in a minority. Furthermore, ICL levels for the GCLASS clusters are very low. Consequently, they do not have a large affect on background measurements made by GALFIT.

\subsubsection{Structural parameters with GALFIT}
\label{GALFIT_parameters}
Rather than opting to use the GALAPAGOS package \citep{Barden2012} to measure the sizes of our cluster and field galaxies, we built our own GALFIT wrapper to have better control over the size determination process. This allowed us to test systematics between the cluster and field data and deal with them individually. This custom built wrapper is run on both the GCLASS and CANDELS-COSMOS mosaics. It is run and tested on the CANDELS-COSMOS mosaic first, to ensure comparable half-light radii to those in \cite{VanderWel2012} are measured. GALFIT is a fitting algorithm that fits two-dimensional analytic functions to light profiles in an image \citep{Peng2002b, Peng2010a}. As was done in \cite{VanderWel2012}, we fit all galaxies with a single-component S\'ersic profile, defined as:
\begin{equation}
    I(r)=I(r_{e})\exp\left[-\kappa\left(\left(\frac{r}{r_{e}}\right)^{\frac{1}{n}}-1\right)\right]
	\label{eq:sersic_profile}
\end{equation}
where $r_{e}$ is the half-light radius. This is the radius within which half of the galaxy's total flux is emitted\footnote{In our study, the half-light radii measurements are {\it not} circularised.}. $n$ is the S\'ersic index and $\kappa$ is an $n$-dependent parameter. $I(r_{e})$ is the intensity at the half-light radius. 

Along with the mosaics and noise maps, GALFIT requires a point-spread function (PSF) as an input. The PSF accounts for the smearing of images due to the resolution limit of WFC3. The PSFs used in this work are stars pre-selected from the GCLASS and CANDELS-COSMOS mosaics under the following criteria: the star must have SExtractor $15 < \mathtt{MAG}$\textunderscore$\mathtt{AUTO} < 19$ and no other sources present near it within a $70~\times~70$ pixel cutout. The presence of other sources is also checked 5 pixels beyond the cutout boundary, to ensure their light does not contribute to the PSF cutout. The reason for this magnitude range is based on preliminary fitting tests, where we found using a star with a magnitude fainter than 19 resulted in poor residuals after subtracting the best-fit model. The cutout dimensions are chosen to encompass the diffraction wings of the star and an adequate amount of sky background. For each galaxy fit, the PSF used from this pre-selected list is the star closest to the galaxy in question in terms of its pixel coordinates on the mosaic. This is done to minimise systematics arising from variations in the PSF across the mosaic.

The GALFIT wrapper will then find all the sources in each cutout and their SExtractor catalogued information. SExtractor values for pixel coordinates, magnitude, $r_{e}$, $n$, axis ratio and position angle are used as initial guesses for every source in the cutout field of view. Every fitting run fits all the sources present in the cutout field-of-view simultaneously. The target galaxy is always the source positioned at the centre of the cutout. We run GALFIT twice for each galaxy fit. The first run keeps all parameters free, with a square fitting region that has a side-length equal to 10\footnote{This cutout size was chosen to ensure that no more than 110 objects were present in a single cutout, since GALFIT can only fit a maximum of 110 objects in one go \citep{VanderWel2012}.} times the SExtractor half-light radius ($\mathtt{FLUX}$\textunderscore$\mathtt{RADIUS}$) of the galaxy in pixels. The purpose of this run is to obtain refined values for the galaxy shape parameters (pixel coordinates, axis ratio and position angle). The second run uses these refined values, but keeps them fixed in the fit, with a square fitting region that has a side-length equal to 15 times the SExtractor half-light radius ($\mathtt{FLUX}$\textunderscore$\mathtt{RADIUS}$) of the galaxy in pixels\footnote{We understand that this cutout size may be too small for the largest of galaxies in the cluster sample, such as the BCGs. We therefore re-fit the BCGs with larger cutouts ($\mathtt{\sim} 100\times100$~kpc), finding that the resulting sizes did not change the results presented in this paper.}. The parameters left free are magnitude, $r_{e}$ and $n$. Although this increases the computational time spent per fit, it vastly improves our size agreement on COSMOS with \cite{VanderWel2012}. This is likely due to the tendency of GALFIT to find the local solution to the parameter estimation. The GALFIT initial values for the shape parameters will be closer to the global solution than the SExtractor values. Hence this two-GALFIT-run approach forces the final fit parameter values closer to the global solution. The level of agreement on COSMOS between the half-light radii measured in this work and \cite{VanderWel2012} is discussed in Appendix~\ref{size_agreement}. In general, we find that there were no systematics present between the two sets of measurements, with a mean offset of $0.28\%$. Our sizes are smaller by this amount. For both GALFIT runs, we keep the sky a free parameter to be fit. We fit every source in the CANDELS-COSMOS and GCLASS mosaics with this procedure, and later filter for the samples required for the mass-size relation study (see Section ~\ref{sample_selection}).

\begin{figure}
	\includegraphics[width=\columnwidth]{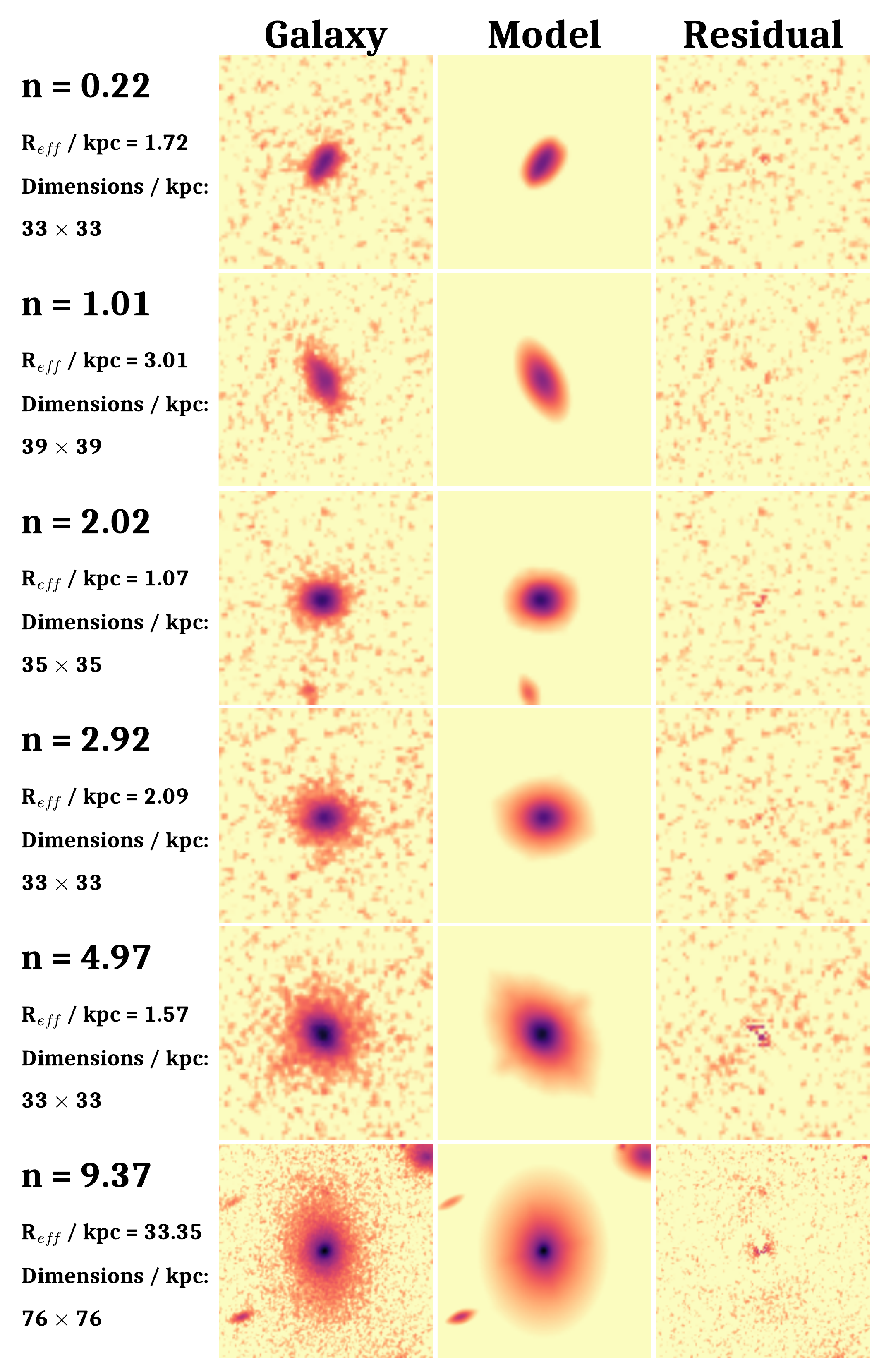}
    \caption{Example S\'ersic profile fits with our GALFIT wrapper for a selection of cluster member galaxies spanning the entire range in S\'ersic index for the cluster sample. First column lists the S\'ersic index and half-light radius measurements for the galaxy, along with the dimensions of the thumbnails. Second, third and fourth columns show the galaxy, model and residual thumbnails for each fit. The colourmap used is logarithmic in scale.}
    \label{fig:sersic_range}
\end{figure}

Approximately $8,000$ galaxies are fit with minimal residuals in the cluster fields-of-view. We show a subset of fits for cluster members in Figure~\ref{fig:sersic_range}, demonstrating the wide range of shape profiles our GALFIT wrapper can cope with.

\subsubsection{Quality check criteria for GALFIT results}

Since the reduced chi-squared results of each GALFIT fit are not necessarily informative on the goodness of fit, we decide to quality check every fit by eye. A GALFIT fit was deemed unusable if one or more of the following criteria are true:
\begin{enumerate}
\item The GALFIT S\'ersic model is more extended than the galaxy, leading to over-subtraction.
\item The position angle of the GALFIT S\'ersic model does not match the position angle of the galaxy.
\item The shape of the GALFIT S\'ersic model is distinctly different to the galaxy.
\item GALFIT fails to create a S\'ersic model.
\end{enumerate}

Any one of these criteria leads to poor residuals and therefore a rejection of fits from the final sample.

\subsection{Sample selection}
\label{sample_selection}
\begin{figure*}
	\includegraphics[width=\textwidth]{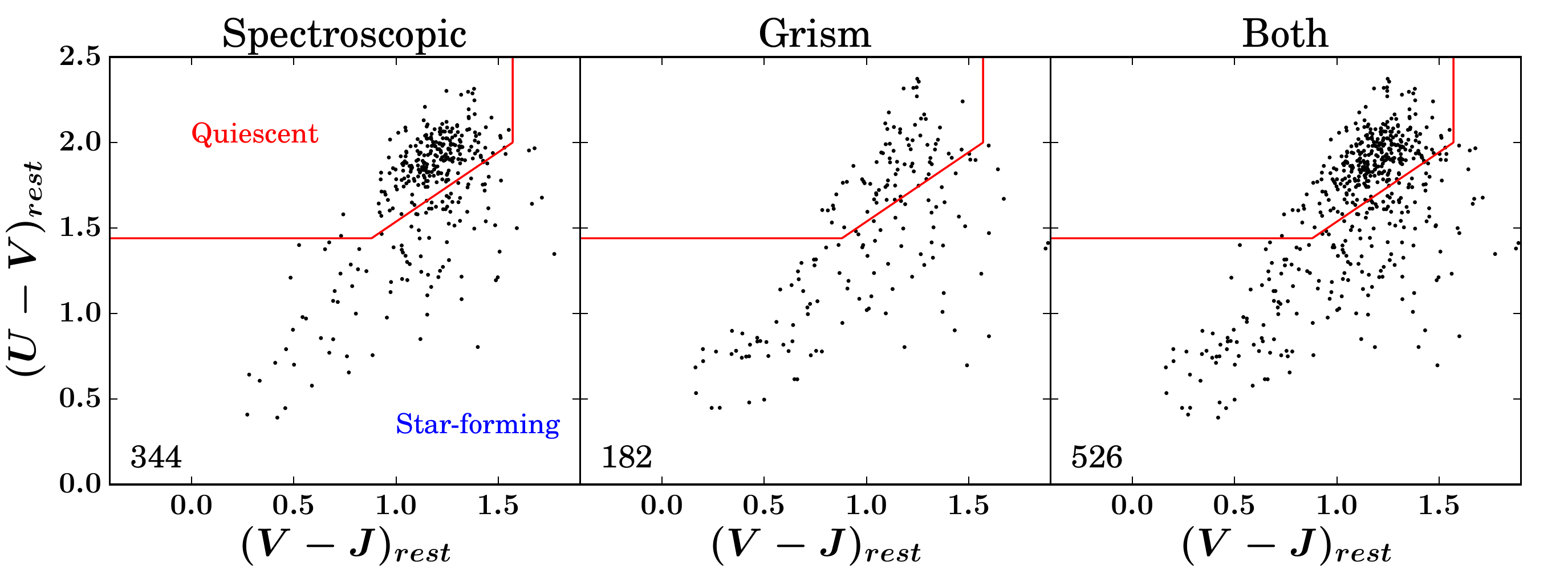}
    \caption{Left: Rest frame $U-V$ versus $V-J$ colour-colour diagram for the spectroscopically confirmed cluster sample. Middle: Rest frame $U-V$ versus $V-J$ colour-colour diagram for the grism-selected cluster sample. Right: Rest frame $U-V$ versus $V-J$ colour-colour diagram for the final cluster sample. Numbers in the bottom left-hand corner of each panel indicate the sample size. The red line shows the dividing line, modified from \protect\cite{Williams2008}, to define the quiescent and star-forming populations. Note: Grism sample includes the 5 cluster members with poor quality GMOS spectra, but good quality grism spectra (see Section~\ref{sample_selection}).}
    \label{fig:clust_col}
\end{figure*}
When selecting samples of galaxies from both high and low density environments, we need to make sure that they are at comparable redshifts and that the galaxies we are selecting from the clusters are indeed cluster members. In this section, we explain in detail how we selected galaxies from both environments so that measurements based on both samples could be compared directly.

As explained throughout Section~\ref{size_determination}, our entire size determination method was first tested on the F160W CANDELS-COSMOS mosaic to ensure it produced comparable sizes to those in \cite{VanderWel2012}. Once this was achieved, we could be confident that our size determination method would produce comparable sizes for the rest of the galaxies in the F160W CANDELS fields. This testing run of our size determination method also allowed us to determine its limits. We find that galaxies in a relevant redshift range to our study \mbox{($0.86 < z < 1.34$)} with a half-light radius measurement from GALFIT of $R_{eff}>50$~kpc or F160W magnitude~$>25$ exhibit the largest disagreements in size with \cite{VanderWel2012}. This divergence in disagreement is driven by increasingly lower signal-to-noise ratio beyond these size and magnitude limits (see Appendix~\ref{size_agreement}). As a result, when we select galaxies from the entirety of the 3D-HST survey, we select those galaxies that are below these size and magnitude thresholds based on measurements made by \cite{VanderWel2012}. We also only choose those measurements that have a flag value of 0 (signifies reliable GALFIT results) according to the \cite{VanderWel2012} catalogue. This gives a sample of $12,501$ field galaxies. We then apply the GCLASS mass completeness limits (see Section~\ref{mass_completeness_limits}) to this sample, reducing it to $3,205$ galaxies.

The same magnitude and size limits for reliable size measurements that were found when testing the size determination method on the F160W CANDELS-COSMOS mosaic are used for the GCLASS measurements as well. Then within these limits for GCLASS, all the spectroscopically confirmed cluster members with GMOS that have good quality spectra and GALFIT residuals are selected as part of the cluster sample. This amounts to 344 galaxies. This number is smaller than the total amount of spectroscopically confirmed cluster members with GMOS (457) because the HST fields-of-view are smaller than those of GMOS. The HST fields-of-view cover $\mathtt{\sim}86\%$ of the spectroscopically confirmed cluster members (see Section~\ref{hst_observations}). Additionally, not all spectroscopically confirmed cluster members obtained reliable size measurements (see Table~\ref{tab:example_table}) due to poor S\'ersic profile fits with GALFIT. The G141 grism data is used in conjunction with all the galaxies that obtained a spectroscopic redshift with GMOS, to determine a selection threshold on grism redshifts, $z_{grism}$ by which additional cluster members can be identified from the grism sample (for more details on this, see Section~\ref{grism_accuracy}). The entire grism sample is first quality-checked by eye. This procedure adds 177 cluster members to the 344 spectroscopically confirmed cluster members. Finally, cluster members that were flagged as ``low confidence" due to poor quality spectroscopic redshifts are checked for good quality grism redshifts consistent with their respective cluster redshifts. This adds 5 more cluster members to the sample. The grism data therefore leads to an addition of 182 cluster members to the 344 spectroscopically confirmed cluster member sample. In total, the final cluster sample amounts to 526 galaxies. A cluster-by-cluster breakdown of the spectroscopic and grism samples can be seen in Table~\ref{tab:example_table}. A summary of this sample selection can be seen in Figure~\ref{fig:clust_col}. Finally, the cluster sample is reduced to 474 galaxies after the GCLASS mass completeness limits are applied to the sample.

\subsubsection{Mass completeness limits}
\label{mass_completeness_limits}
The GCLASS mass completeness limits are set to match the grism spectroscopic completeness limits of the final cluster member sample. Every galaxy in the final cluster member sample (reliable grism spectra, F140W magnitude~$<25$ and $R_{eff}<50$~kpc) has a good quality grism spectrum from which reliable grism redshifts are calculated, confirming their cluster membership (see Section~\ref{grism_accuracy} for more details). The GCLASS mass completeness limits are calculated for the star-forming and quiescent cluster galaxies in the final cluster member sample separately, using the following method: the luminosity distances and absolute magnitudes of all cluster galaxies are calculated. Stellar mass-to-light ratios for all cluster galaxies are then calculated. The cluster galaxy with the highest mass-to-light ratio is found. This galaxy is the spectroscopically confirmed (either spectroscopic or grism) cluster galaxy for which the most stellar mass is measured for the least amount of flux. Then, the faintest galaxy with a spectroscopic or grism redshift in the sample is found. We then calculate what stellar mass this cluster galaxy would have, if it had a mass-to-light ratio corresponding to the highest value measured in the sample. This provides us with the lowest stellar mass for which a reliable grism spectrum can be obtained. We find mass completeness limits of log$(M_{*}/M_{\odot})=9.96$ and log$(M_{*}/M_{\odot})=9.60$ for quiescent and star-forming cluster galaxies, respectively. These mass completeness limits are comparable to those calculated by \cite{VanderBurg2013} for the GCLASS clusters using ground-based {\it K}-band data. The limits used in our study are deeper due to the increased depth of the F140W imaging in comparison to the ground-based {\it K}-band data.
 
\begin{figure}
	\includegraphics[width=\columnwidth]{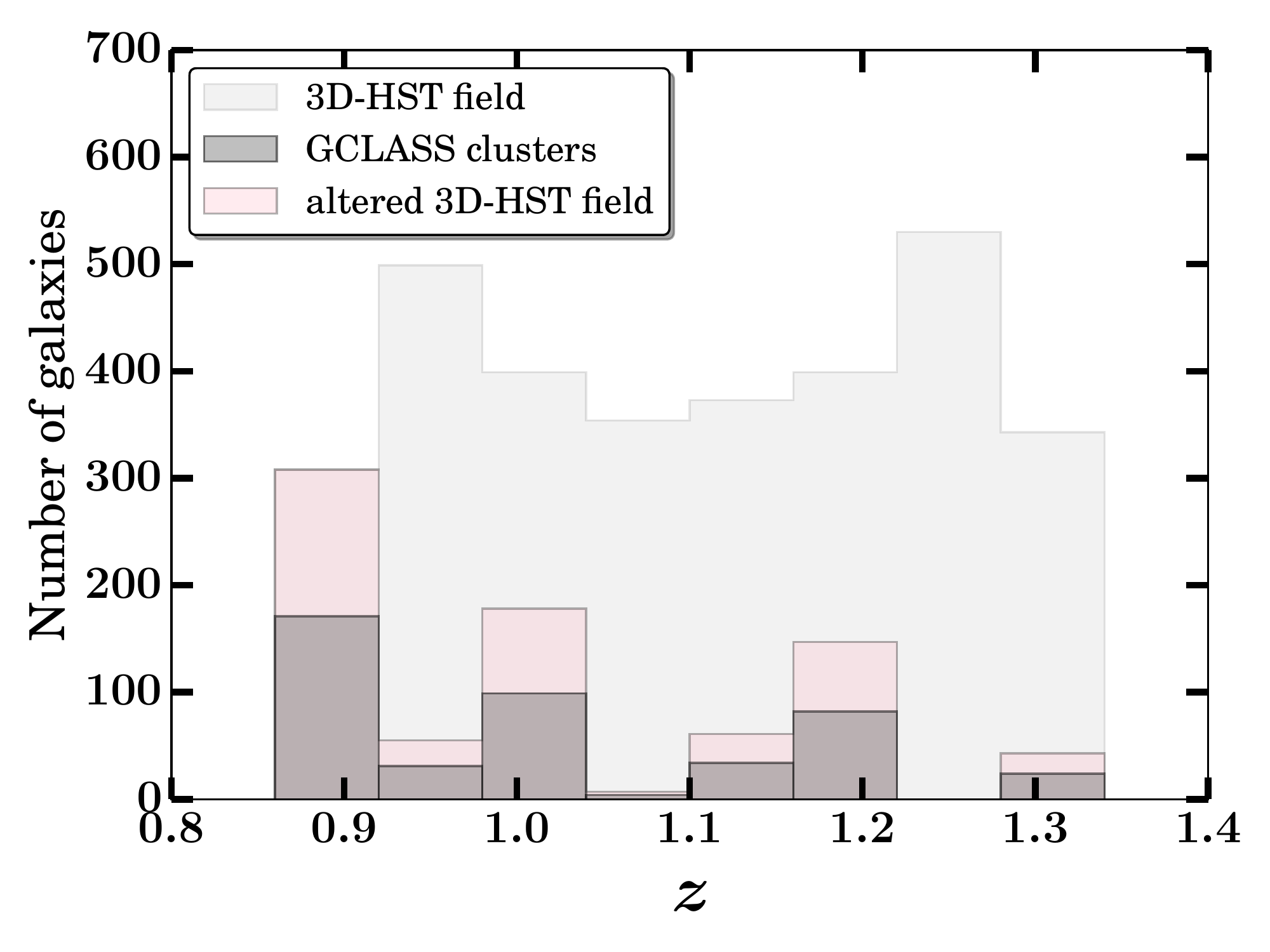}
    \caption{Alteration of the field redshift distribution, such that it follows the shape of the cluster redshift distribution. The original field sample (light grey histogram) contains more galaxies in every redshift bin compared to the cluster sample (dark grey histogram). The bin with the largest number of galaxies in the cluster sample (first bin) is used to work out its height ratio with the corresponding bin in the field. The heights of all the other cluster bins are scaled up using this ratio, leading to the distribution shown in pink.}
    \label{fig:bootstrap}
\end{figure}

\subsubsection{Redshift distribution matching}
\label{redshift_distribution}
After the mass completeness limits have been applied to the field sample, we draw 1000 random field samples in a bootstrap-like method that follow the same redshift distribution as the cluster sample. This is done in the following way: histograms of the cluster and field sample redshifts are plotted with the same redshift bins. Since there are more galaxies in the field sample, each bin in the field sample contains more galaxies than the same bin in the cluster sample. We take the bin with the largest number of galaxies in the cluster sample and calculate its normalisation with the corresponding field sample bin. This provides us with a factor with which we can multiply the height of all the cluster sample bins, and this establishes how many field galaxies should populate each bin. As a result, a distribution of field galaxy redshifts is obtained that matches the shape of the cluster distribution exactly. Figure~\ref{fig:bootstrap} more clearly illustrates this process. Field galaxies are then randomly selected 1000 times within each of these redshift bins. The total number of field galaxies selected per bin corresponds to the height of each pink redshift bin shown in Figure~\ref{fig:bootstrap}. The result is 1000 different field samples, each containing 799 galaxies with the same mean redshift as the cluster sample. This alteration in the field redshift distribution is needed because when we select 3D-HST CANDELS galaxies in the redshift range $0.86<z<1.34$, the median $z$ is 1.09, however, the median redshift of the cluster distribution is $z=1.00$. This difference is large enough to introduce a systematic in our size offset measurements in Section~\ref{mass_size_sec} of order our uncertainties (up to $0.03$ dex for quiescent galaxies and up to $0.01$ dex for star-forming galaxies) \citep{VanderWel2014}. Consequently, many field samples are required to measure reliable size offsets with the cluster sample since the mass-size distributions can vary between field samples. This leads to variations in the field relations that are fit, from which size offsets are measured (see Section~\ref{mass_size_sec} for more details).

\begin{figure*}	\includegraphics[width=\textwidth]{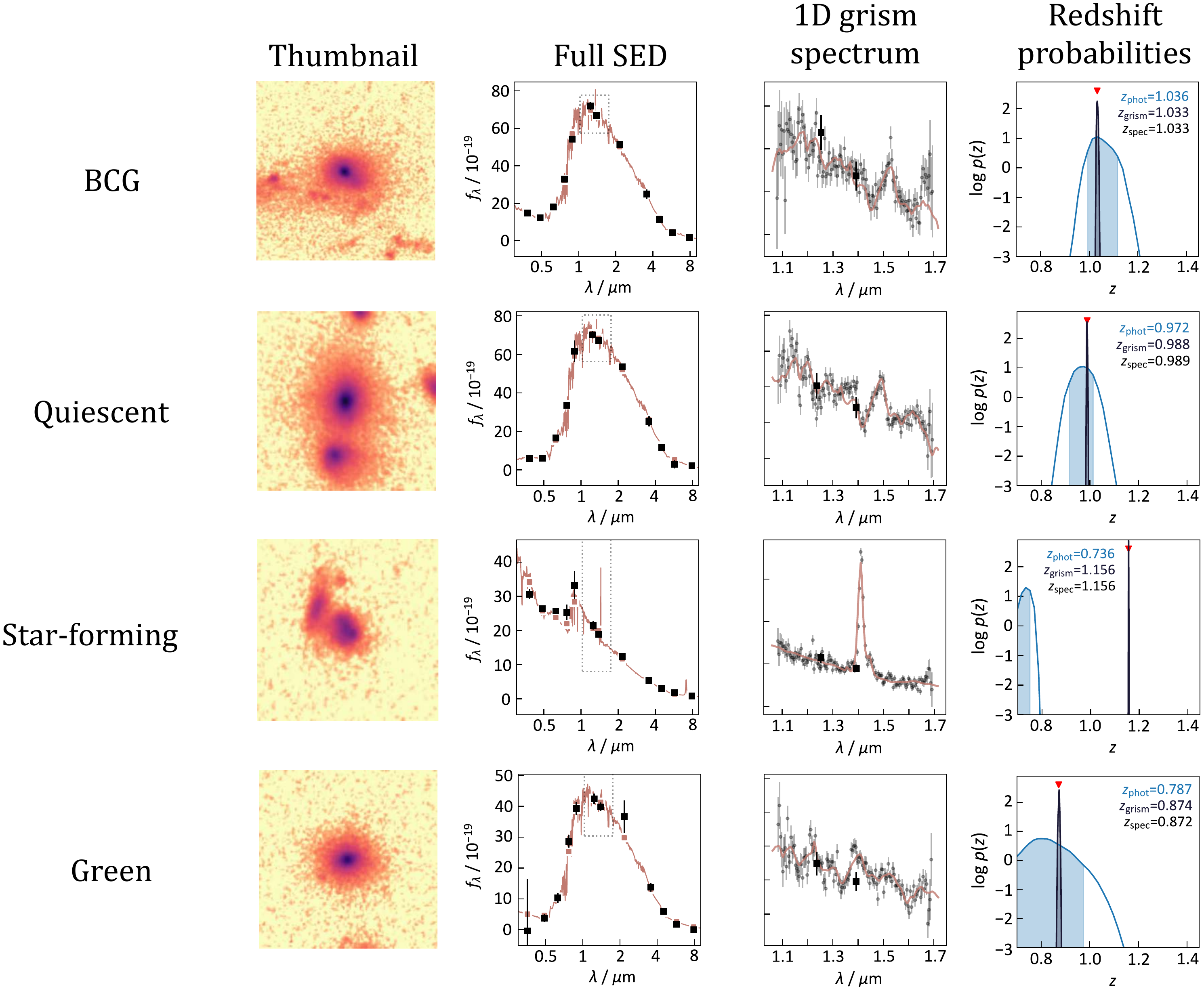}
    \caption{First column: cutouts from the GCLASS HST F140W mosaics of an example BCG, quiescent, star-forming and green cluster galaxy that have been spectroscopically confirmed. Colourmap is logarithmic. Second column: Full SEDs of the galaxies from the 12-band photometry. The dotted grey rectangle highlights the region in which the grism spectrum lies. Third column: A zoom-in of the regions highlighted with dotted grey rectangles in the second column. Data points from the raw-one dimensional grism spectra are also shown in grey. Black data points in both the second and third columns are from the 12-band photometry. Fourth column: redshift probability distributions. Photometric redshift distributions are shown in blue, with the shaded region representing the $2\sigma$ range. Grism redshift distributions are shown in black and red triangles show the location of the spectroscopic redshift.}
    \label{fig:grism_example_spec}
\end{figure*}

\subsubsection{Rest-frame colours}
\label{colors}
EAZY \citep{Brammer2008} is used to interpolate the input SED to obtain the $U-V$ and $V-J$ rest-frame colours for each galaxy \citep{VanderBurg2013}.
This $UVJ$ colour separation for star-forming and quiescent populations has been shown to be a powerful technique in separating the two populations, even if the former is reddened by dust extinction \citep{Wuyts2007, Williams2008, Patel2012}. This method is also used on the field sample. The dividing line used for the field sample is an interpolation of the lines defined in \cite{Williams2008} for \mbox{$0.5 < z < 1.0$} and  \mbox{$1.0 < z < 2.0$}. Quiescent field galaxies are defined as

\begin{equation}
  (U-V)_{rest} > 1.3,~~~(V-J)_{rest} < 1.6
	\label{eq:field_line_1}
\end{equation}
and
\begin{equation}
  (U-V)_{rest} > 0.88~(V-J)_{rest}+0.54
	\label{eq:field_line_2}
\end{equation}
 We modify this dividing line for GCLASS to fit our rest-frame colour distribution. This is the red line shown in Figure~\ref{fig:clust_col}, with quiescent cluster galaxies defined as 

\begin{equation}
  (U-V)_{rest} > 1.44,~~~(V-J)_{rest} < 1.57
	\label{eq:cluster_line_1}
\end{equation}
and
\begin{equation}
  (U-V)_{rest} > 0.81~(V-J)_{rest}+0.73
	\label{eq:cluster_line_2}
\end{equation}

An example quiescent, star-forming, green (close to where Equation~\ref{eq:cluster_line_2} becomes an equality) and Brightest Cluster Galaxy (BCG) from the GCLASS spectroscopic sample are shown in Figure~\ref{fig:grism_example_spec} with their accompanying grism data and measurements. The second column shows the full SEDs from the 12-band photometry. The grey dotted rectangles in these plots show the location of the grism spectra with respect to the SEDs. A zoom-in of these regions is seen in the third column, with data points from the raw one-dimensional grism spectra. The final column shows the redshift probability distributions based on fitting the two-dimensional grism spectra (see Figure~\ref{fig:grism_eg}) and the 12-band photometry simultaneously (see Section~\ref{cluster_sample} for more details). This is done using a modified version of the EAZY code \citep{Brammer2008}. The extensive photometric coverage for GCLASS produces improved fits to the SEDs and grism spectra. The level of improvement is best seen in the almost one-to-one agreement with the spectroscopic and grism redshifts in the redshift probability distributions. What is perhaps more striking is the improvement in redshift estimates between the photometrically derived redshifts, $z_{phot}$ and the grism redshifts, $z_{grism}$. In Section~\ref{grism_accuracy}, we will demonstrate how this improved level of accuracy on redshift estimates is utilised to increase our cluster membership sample.

\section{The precision \& accuracy of grism redshifts}
\label{grism_accuracy}

\begin{figure*}
	\includegraphics[width=\textwidth]{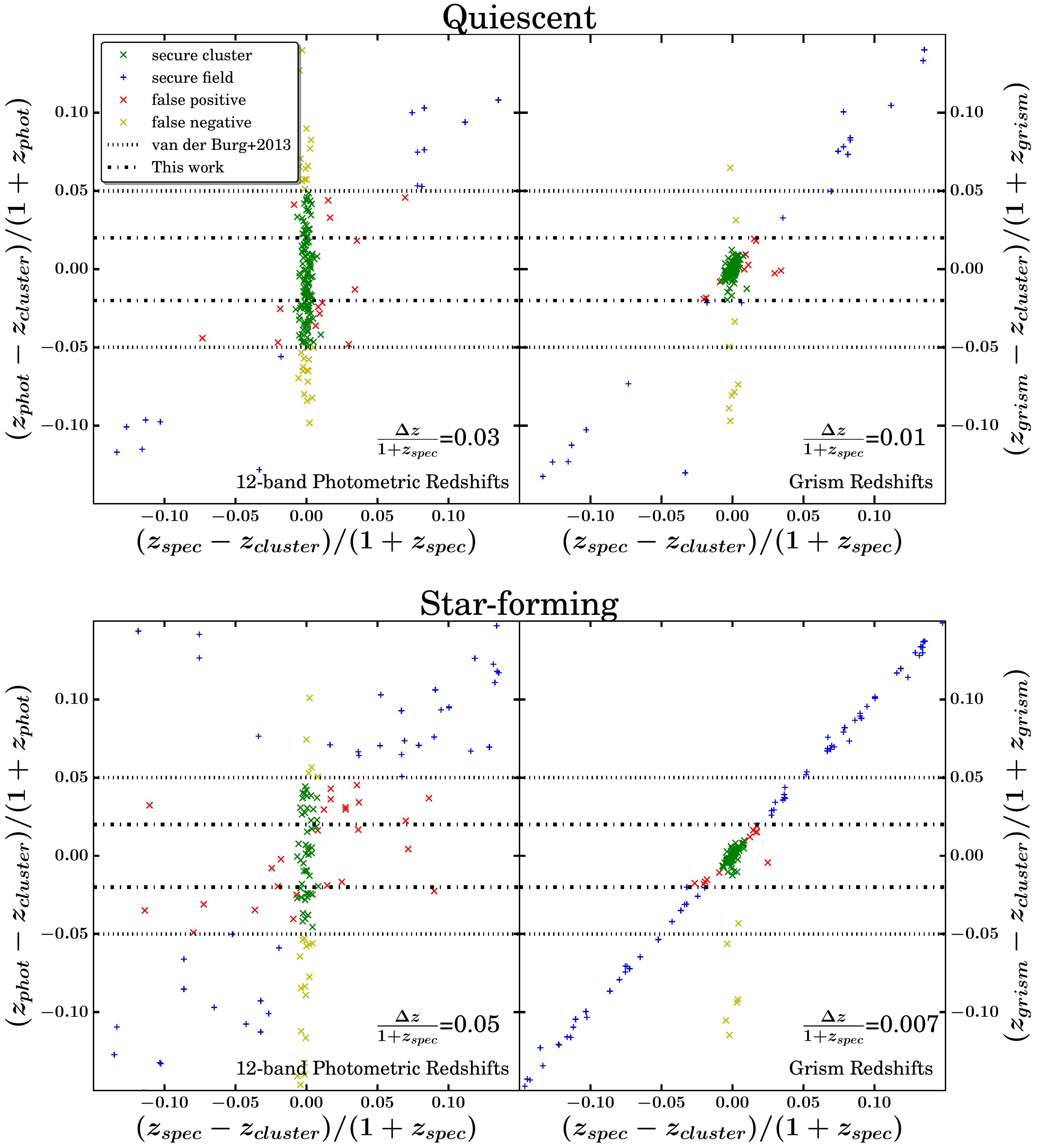}
    \caption{Top row: comparison of spectroscopic redshift measurements ($z_{spec}$) with photometric (left) and grism (right) redshift measurements ($z_{phot}$ and $z_{grism}$) for quiescent galaxies in the HST fields of view with reliable $z_{spec}$ and $z_{grism}$ measurements. Bottom row: Same as the first row, but for star-forming galaxies. All comparisons are done relative to the spectroscopic redshift of the cluster centre (in most cases the BCG), $z_{cluster}$. Dotted black lines show the photometric redshift selection threshold from \protect\cite{VanderBurg2013} used to select cluster members. The dot-dashed black lines show the grism redshift selection threshold chosen for this work after comparing $z_{spec}$ and $z_{grism}$ measurements for this sample (second column). $\Delta z = |z_{phot} - z_{spec}|$ for the first column and $|z_{grism} - z_{spec}|$ for the second column. Values for $\frac{\Delta z}{1+z_{spec}}$ shown are the averages for the entire sample in that respective panel.}
    \label{fig:grism_spec_phot3}
\end{figure*}

We take all the galaxies in the GCLASS HST fields-of-view that have reliable spectroscopic and grism redshifts and compare their redshift measurements from photometry ($z_{phot}$), spectroscopy ($z_{spec}$) and grism spectroscopy ($z_{grism}$) to their respective cluster redshifts, $z_{cluster}$. We do this to determine a selection threshold by which cluster members can be identified from the sample of galaxies that did not obtain a spectroscopic redshift with GMOS, but did obtain a grism redshift. To do this, we use the following terminology that was first introduced in \cite{VanderBurg2013}. A ``false positive" is a galaxy that is not part of the cluster by virtue of its $z_{spec}$, but has a $z_{phot}$ or $z_{grism}$ consistent with the cluster redshift. A ``false negative" is a galaxy that belongs to the cluster due to its $z_{spec}$ measurement, but has a $z_{phot}$ or $z_{grism}$ placing it outside the cluster. ``secure cluster" galaxies are those which are classified as being part of the cluster by virtue of both their $z_{spec}$ and $z_{phot}$ or $z_{grism}$. ``secure field" galaxies are consistent with being outside the cluster due to their $z_{spec}$ and $z_{phot}$ or $z_{grism}$ measurements. A clearer representation of this can be seen in Figure~\ref{fig:grism_spec_phot3}. The top row shows the comparison between $z_{phot}$ and $z_{grism}$ measurements for quiescent galaxies with good quality $z_{spec}$ measurements. The bottom row shows the same, but for star-forming galaxies. The quiescent and star-forming separation is done using the $UVJ$ selection criteria discussed in Section~\ref{colors}. What is immediately striking is the tight clustering of green crosses in the central regions of the $z_{grism}$ vs. $z_{spec}$ comparison plots. These are all grism-redshift selected cluster members with a spectroscopic redshift to support the reliability of their cluster membership. Another noticeable feature is all the star-forming galaxies that lie on an almost one-to-one correspondence line in the $z_{grism}$ vs. $z_{spec}$ comparison. This is another representation of the excellent level of agreement between $z_{spec}$ and $z_{grism}$ measurements, attributed both to the H$\alpha$ emission line that is prevalent in star-forming galaxy grism spectra (see Figure~\ref{fig:grism_example_spec}) and the 12-band GCLASS photometry.

The black horizontal dotted lines represent the cluster membership selection threshold used in \cite{VanderBurg2013} based on a $z_{phot}$ and $z_{spec}$ comparison. When comparing the plots in the first and second columns, it is clear that there is no obvious selection threshold in the $z_{phot}$ measurements by which cluster members can be reliably selected. The \cite{VanderBurg2013} selection threshold of $-0.05 < \Delta z < 0.05$ was selected due to its convenience in producing a similar number of false positives and false negatives. The higher precision of the grism redshift measurements allow us to choose a selection threshold that is much narrower, lying at $-0.02 < \Delta z < 0.02$ (black dot-dashed line). This increases the number of cluster members per cluster by an average of $53\%$ of the spectroscopic sample, due to galaxies without a $z_{spec}$ measurement, but with a reliable $z_{grism}$ measurement placing them inside their respective cluster (see Table~\ref{tab:example_table}). Our selection threshold has more than a factor of 2 improvement over the one used in the photometric cluster membership selection in \cite{VanderBurg2013}. We have also listed the average difference between the $z_{phot}$ and $z_{grism}$ measurements with the $z_{spec}$ measurements for each sample in Figure~\ref{fig:grism_spec_phot3}. The precision of photometric and grism redshifts compared to spectroscopic redshifts in kms$^{-1}$ is $8400$ and $2000$ kms$^{-1}$ respectively: a factor of 4 improvement of grism redshifts over photometric redshifts. A factor of 3 and 7 improvement is seen for quiescent and star-forming galaxies respectively. The fraction of false negatives and false positives to secure cluster members falls from $28\%$ and $23\%$ to $8\%$ and $10\%$, a factor of $\mathtt{\sim}3$ improvement in contamination. Due to these low contamination rates, we do not perform corrections on our grism-selected cluster membership sample like those that were done on the photometric-selected cluster membership sample in \cite{VanderBurg2013}.

It must be stressed that this impressive improvement in precision and accuracy is due to the 12-band photometry GCLASS possesses. Both photometric and grism redshift measurements rely on the SEDs since they are used in both fitting processes. Therefore greater SED quality results in increased measurement precision and accuracy for both $z_{phot}$ and $z_{grism}$ measurements. A larger number of photometric bands --- especially in the optical and near infra-red --- for a given wavelength range improve the quality of the resulting SEDs from which photometric redshifts, grism redshifts and stellar masses are derived. \cite{Bezanson2016} studied the accuracy of grism redshifts with respect to photometric redshifts for the 3D-HST survey. They found similar values for the scatter in $z_{phot}$ vs. $z_{spec}$ and $z_{phot}$ vs. $z_{grism}$ ($0.0159\pm0.0005$ and $0.0154\pm0.0005$), highlighting that grism and spectroscopic redshifts are of equal quality relative to photometric redshifts. We also find similar values ($0.06\pm0.06$ and $0.07\pm0.07$) with overall values $\mathtt{\sim}4$ times higher than those of 3D-HST. Given that our methods for deriving photometric redshifts and stellar masses are near-to identical (see Section~\ref{photometric_redshifts}) and that our photometry spans the same wavelength range as for 3D-HST, the one thing we do not have is a comparable number of photometric bands and signal-to-noise ratio. The five 3D-HST fields (see Section~\ref{field_sample}) have photometry spanning between 18 (UDS) and 44 (COSMOS) bands, going deeper by 1--3 magnitudes when comparing common photometric bands between GCLASS (Appendix A of \citealt{VanderBurg2013}) and 3D-HST \citep{Skelton2014a}. 3D-HST's superior photometry allowed for better photometric and grism redshift measurements compared to those for GCLASS.

Nevertheless, the GCLASS HST data has demonstrated that grism redshift measurements derived from good quality SEDs can be used to select samples of galaxies for cluster membership that are $\mathtt{\sim}90\%$ pure. Space-based grism slitless spectroscopy -- used in conjunction with photometry spanning multiple bands -- therefore provides a unique alternative to conventional spectroscopic surveys, especially when large samples are required. JWST, WFIRST and {\it Euclid} will all have grism capability, with part of their science goals dedicated to studying high-redshift galaxies \citep{Gardner2006,Report2011,Green2012}. This technical result demonstrates that grisms on future space-based telescopes will be able to provide large samples of cluster galaxies with high reliability.

\section{The cluster vs. field stellar mass-size relation at $\lowercase{z}~\mathtt{\sim}~1$}
\label{mass_size_sec}

\begin{figure*}
	\includegraphics[width=\textwidth]{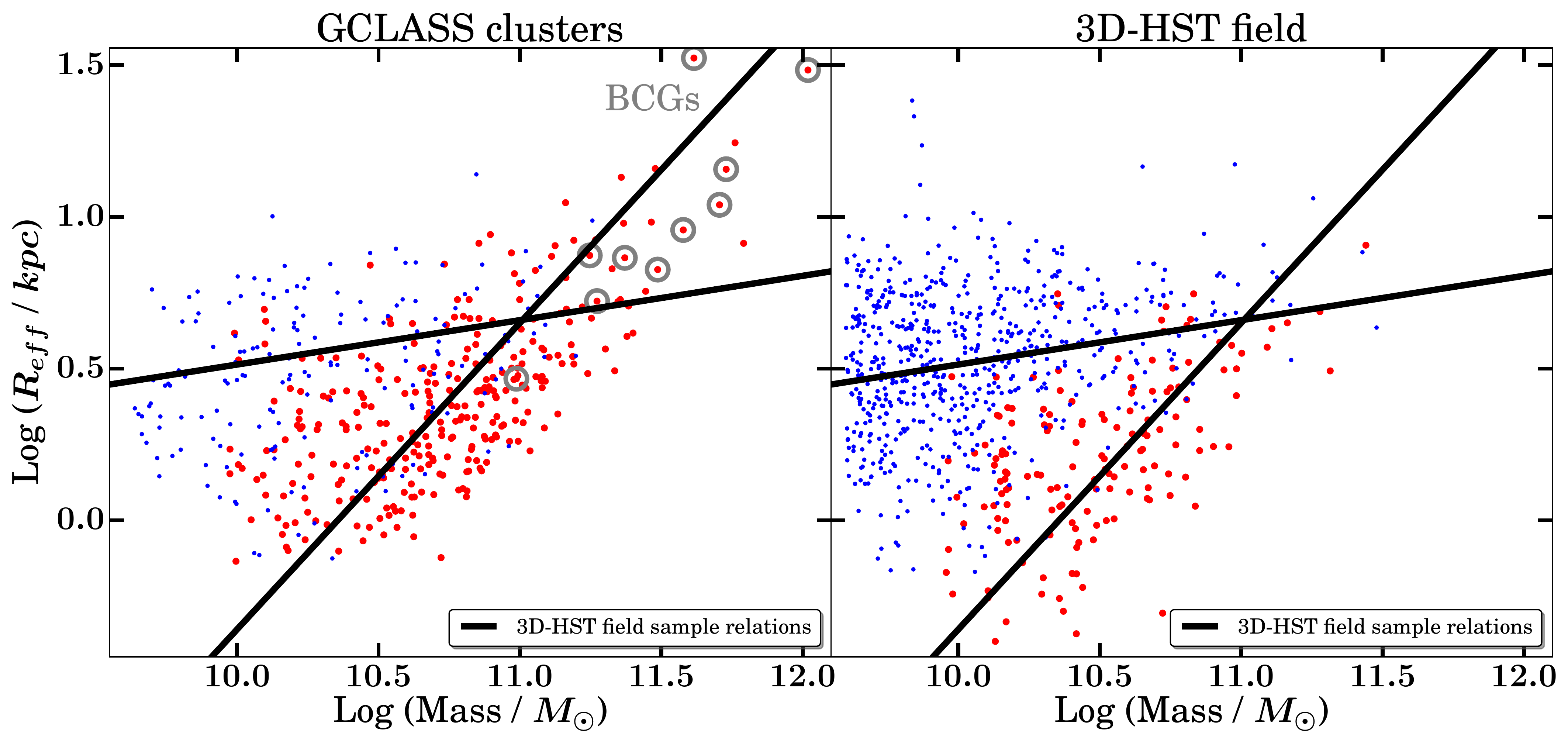}
    \caption{Stellar mass-size relation for the GCLASS clusters (left panel) and for one of the 3D-HST field samples (right). Smaller blue (larger red) circles show star-forming (quiescent) galaxies within the mass completeness limits of GCLASS. $R_{eff}$ is the half-light radius in kpc. GCLASS Brightest Cluster Galaxies (BCGs) are circled in grey. Solid black lines show the field relations at $z~\mathtt{\sim}~1$ calculated using results from \protect\cite{VanderWel2012} (see text). }
    \label{fig:mass_size_v2}
\end{figure*}

\begin{figure}
	\includegraphics[width=\columnwidth]{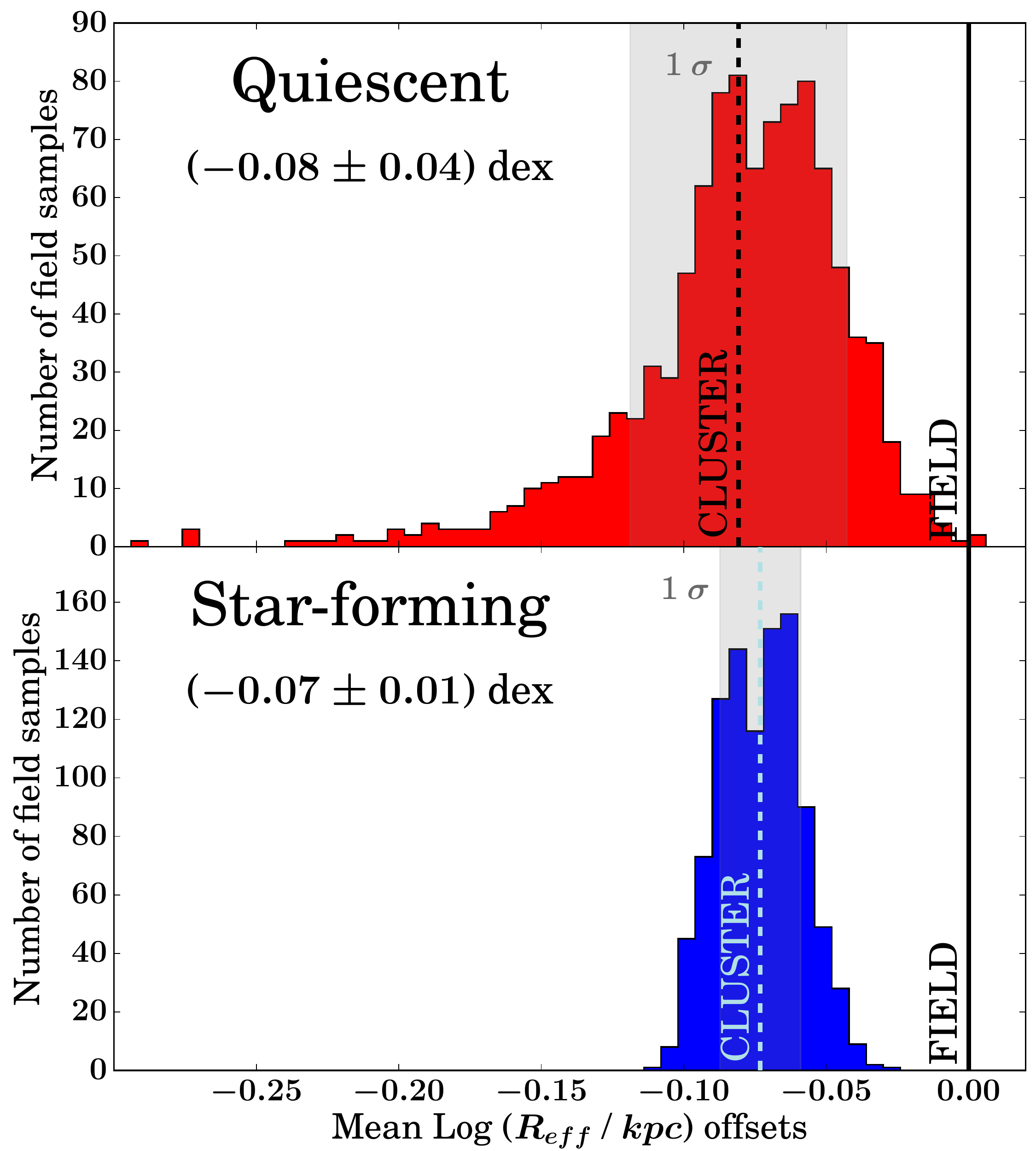}
    \caption{Histograms of the differences between the mean size offsets of cluster and field galaxies in log space (mean cluster size offset - mean field size offset). Cluster and field size offsets are measured relative to the 1000 possible field mass-size relations for the 1000 different field samples. This is done for quiescent and star-forming galaxies separately. Solid black vertical lines show the location of no difference in average size between field and cluster galaxies at fixed stellar mass. Dashed vertical lines show the mean of all the differences measured in the mean size offsets of cluster and field galaxies. Grey shaded regions show one standard deviation in the mean size offset differences. On average, quiescent and star-forming cluster galaxies are smaller than their field counterparts by ($0.08\pm0.04$) dex and ($0.07\pm0.01$) dex respectively.}
    \label{fig:offset_hists}
\end{figure}
The high quality grism redshift measurements for GCLASS provide us with the largest spectroscopically-confirmed sample of star-forming and quiescent cluster galaxies at $z~\mathtt{\sim}~1$ for which the stellar mass-size relation can be measured. By first testing the reliability of our size determination method using published results for the CANDELS-COSMOS F160W mosaic (see Appendix~\ref{size_agreement}), we have ensured our measurements for the cluster sample will be robust. Figure~\ref{fig:mass_size_v2} shows the mass-size relation for the 10 GCLASS clusters in the left panel, with the BCGs\footnote{The BCGs were identified as the brightest cluster member in the observer-frame $K_{s}$ band \citep{Lidman2012}.} circled in grey. The solid black field lines were calculated using the 1000 field samples (see Section~\ref{redshift_distribution}) in the following way: The stellar mass (corrected for the difference in flux measurements, see Section~\ref{photometric_redshifts}) and half-light radii measurements with uncertainties from \cite{VanderWel2012} for the star-forming and quiescent field galaxies in our 1000 field samples are used to fit for straight lines in log space using the Bayesian method from \cite{Kelly2007}. The star-forming and quiescent field relations are fit separately for slope, intercept and intrinsic scatter for each of the 1000 samples. The mean slope and intercept value from these 1000 fits are used to calculate the final field relations, shown in Figure~\ref{fig:mass_size_v2}. We choose this method of line-fitting because we find that using simple orthogonal distance regression does not produce field relations that pass through the mean of the mass-size measurements used to fit the relations. It is therefore apparent that a fitting method which also fits for the intrinsic scatter is essential to producing realistic field relations. When fitting, we only use star-forming and quiescent field galaxies that meet the GCLASS mass completeness limits and are within the stellar mass thresholds used in the fitting process described in \cite{VanderWel2014} (star-forming galaxies with $M_{*} > 3\times10^{9}~M_{\odot}$ and quiescent galaxies with $M_{*} > 2\times10^{10}~M_{\odot}$). In an attempt to follow the fitting method in \cite{VanderWel2014} as closely as possible, we add an uncertainty of $0.15$ dex in the \mbox{Log~(Stellar Mass~/~$M_{\odot}$)} measurements. For comparison, in the right panel of Figure~\ref{fig:mass_size_v2}, we plot the stellar mass-size relation for one of the 1000 field samples for which the resulting field relation fits were close to the final field relations.

We then compare the stellar mass-size measurements for both the field and cluster samples more rigorously. We measure the size offsets in log space from each of the 1000 field relations for each of the 1000 field samples. We also measure the size offsets of the cluster sample from each of these 1000 field relations. We then take the mean of each size offset distribution. The difference between the field and cluster mean size offsets is calculated for each of the 1000 field samples. The distribution of these differences in the mean size offsets in log space for quiescent and star-forming galaxies is shown in Figure~\ref{fig:offset_hists}. The reason we do not select one field sample and measure size offsets from a single field relation is because the field mass-size distribution varies between the 1000 field samples. As a result, the calculated field relations vary. This is particularly important for calculating the quiescent field relations. There are fewer quiescent galaxies compared to star-forming galaxies in the field samples (evident in Figure~\ref{fig:mass_size_v2}). As a result, fewer galaxies are used to fit for the quiescent field relations than for the star-forming field relations. This leads to wider variations in the slope and intercept values for the 1000 quiescent field relations compared to the variations found for the 1000 star-forming field relations. When we take the mean of all individual average size offset differences measured between cluster and field, we find that quiescent and star-forming cluster galaxies are smaller than their field counterparts by ($0.08\pm0.04$) dex and ($0.07\pm0.01$) dex respectively. The result for quiescent cluster galaxies does not change significantly if we remove the BCGs from the cluster sample. The magnitude of these size offsets, particularly in the case of quiescent galaxies, is exactly what we would expect from the minor-mergers hypothesis of size growth advocated by \cite{VanderWel2014} and many others (e.g. \citealt{Bezanson2009,Hopkins2009c,VanDokkum2010}). 

In the minor-mergers hypothesis of size growth, galaxies are able to grow disproportionately more in size compared to stellar mass with decreasing redshift. This is because minor mergers cause mass to be preferentially deposited at larger radii. Such growth would be able to increase the intercept of the mass-size relation without significantly affecting the slope. This is seen in the field for 3D-HST between $0<z<3$ in \cite{VanderWel2014}. In order to test whether it is indeed minor mergers driving the evolution in the intercept, we need a population of galaxies that we know are virtually unable to grow via minor mergers, and compare their stellar mass-size relation to 3D-HST at fixed redshift. This population is represented by our cluster galaxies. The high velocity dispersions in clusters suppress mergers from occurring, consequently suppressing size growth via minor mergers. If minor mergers are truly responsible for the majority of the disproportionate size growth observed in the field, we should find that the sizes of cluster galaxies are offset to smaller sizes at fixed redshift. The magnitude of this offset needs to be equivalent to the expected evolution in the intercept of the field mass-size relation between the redshift at which the current cluster galaxies fell into their clusters and $z~\mathtt{\sim}~1$. If the magnitude of the size offset is significantly smaller than this value, we cannot be sure that minor mergers are responsible for the majority of the size growth observed in the field.

A cluster with the average mass of a GCLASS cluster ($4.2\times10^{14}~M_{\odot}$; see Table~\ref{tab:example_table}) at $z~\mathtt{\sim}~1$ is expected to accrete most of its mass by approximately $1.2\leqslant~z~\leqslant1.5$ \citep{Fakhouri2010,VanderBurg2014a}\footnote{This is found by looking at the halo mass a cluster with the average $M_{200}$ of a GCLASS cluster at $z~\mathtt{\sim}~1$ would reach by $z~\mathtt{\sim}~0$ in the Millenium Simulations. The redshift at which approximately $>50\%$ of the mass is assembled is read-off from the top panel of Figure 6 in \cite{Fakhouri2010}.}. Under the assumption that most of the current GCLASS cluster members fell into their clusters from the field at these redshifts -- and had their size growth suppressed since then -- they should exhibit sizes that follow the field stellar mass-size relations at these redshifts. We use the preferred parameterisations for the evolution in the intercept of the field mass-size relation found in \citealt{VanderWel2014} ($R_{eff}~/$~kpc~$=~4.3~h(z)^{-1.29}$ for quiescent galaxies and $R_{eff}~/$~kpc~$=~7.8~h(z)^{-0.66}$ for star-forming galaxies) to check the expected size offset in the field between $z~\mathtt{\sim}~1$ and $1.2\leqslant~z~\leqslant1.5$. We find size offset ranges of $0.06-0.2$ dex and $0.03-0.08$ dex for quiescent and star-forming galaxies respectively. Our results of ($0.08\pm0.04$) dex and ($0.07\pm0.01$) dex for quiescent and star-forming galaxies are therefore consistent with these allowed ranges, supporting the possibility that minor mergers are the dominant route for galaxy size growth.

 Previous work on measuring the difference between the field and cluster stellar mass-size relations at fixed redshift has consisted of a diverse range of results. At low redshifts ($z<0.2$), \cite{Cebrian2014} is the only study which finds smaller sizes in both early-type and late-type galaxies in the cluster environment. They find that early-type and late-type cluster galaxies are $\mathtt{\sim}3.5\%$ and $\mathtt{\sim}7.5\%$ smaller than their field counterparts respectively. These percentages are approximately what is expected at these redshifts if we use the same parameterisations from \cite{VanderWel2014} and assume that most of the current cluster members in the clusters used in \cite{Cebrian2014} fell into their clusters at a redshift that corresponds to $\mathtt{\sim}~1$~Gyr prior to observation. However, \cite{Weinmann2009} and \cite{Maltby2010} find no difference in early-type galaxies with environment. At a similar redshift to our work, \cite{Delaye2014a} studied the mass-size relation for massive early-type galaxies in 9 clusters compared to a field sample selected from a variety of datasets. They found no difference in the median sizes with environment, but a skew towards larger sizes in clusters by $30-40\%$. This is a significantly different result to ours, where we find early-type galaxies are $\mathtt{\sim}~20\%$ {\it smaller} in clusters compared to the field, at fixed stellar mass. At a slightly higher redshift of $z\mathtt{\sim}1.3$, \cite{Raichoor2012} found early-type galaxies to be $30-50\%$ smaller in clusters. This result is also consistent with what is expected at this redshift using the intercept evolution parameterisation for quiescent galaxies from \cite{VanderWel2014} and assuming the current cluster members fell into their clusters at a redshift corresponding to $\mathtt{\sim}~1$~Gyr prior to observation. We postulate that these conflicting results in the literature might be due to a combination of small sample sizes at high redshifts and the difficulty in minimising systematics between field and cluster samples.

We have shown that when a careful treatment of systematics between datasets is carried out, there is a difference between the field and cluster stellar mass-size relations at $z~\mathtt{\sim}~1$. The magnitude of this difference is consistent with the expected size growth from minor mergers in the field for the likely duration the current cluster galaxies at $z~\mathtt{\sim}~1$ have been in their clusters.

It should be mentioned that size growth in the quiescent field population is also thought to be due to newly-quenched field galaxies joining the quiescent field population at later times \citep{Carollo2013}. Galaxies that quench later are larger than galaxies that quenched earlier. Therefore, if a large fraction of the quiescent field population consists of newly-quenched galaxies, quiescent field galaxies would be on average larger than quiescent cluster galaxies at fixed redshift and stellar mass. Nevertheless, while this does play a role in the average size growth of the quiescent field population, it is not thought to be the only explanation, with merging thought to be the other important contribution (e.g. \citealt{VanDerWel2009}).

Our suggestion that the lack of minor mergers in the cluster environment inhibits size growth would consequently lead to a significant difference between the cluster and field environment at low redshifts. In Section~\ref{BCG_growth} we will discuss the implications of this in more detail.

In the next section, we will investigate whether the differences in the cluster and field mass-size relations are due to morphological differences.

\section{Morphology and the cluster vs. field stellar mass-size relation at $\lowercase{z}~\mathtt{\sim}~1$}
\label{morphology}
\begin{figure*}
	\includegraphics[width=\textwidth]{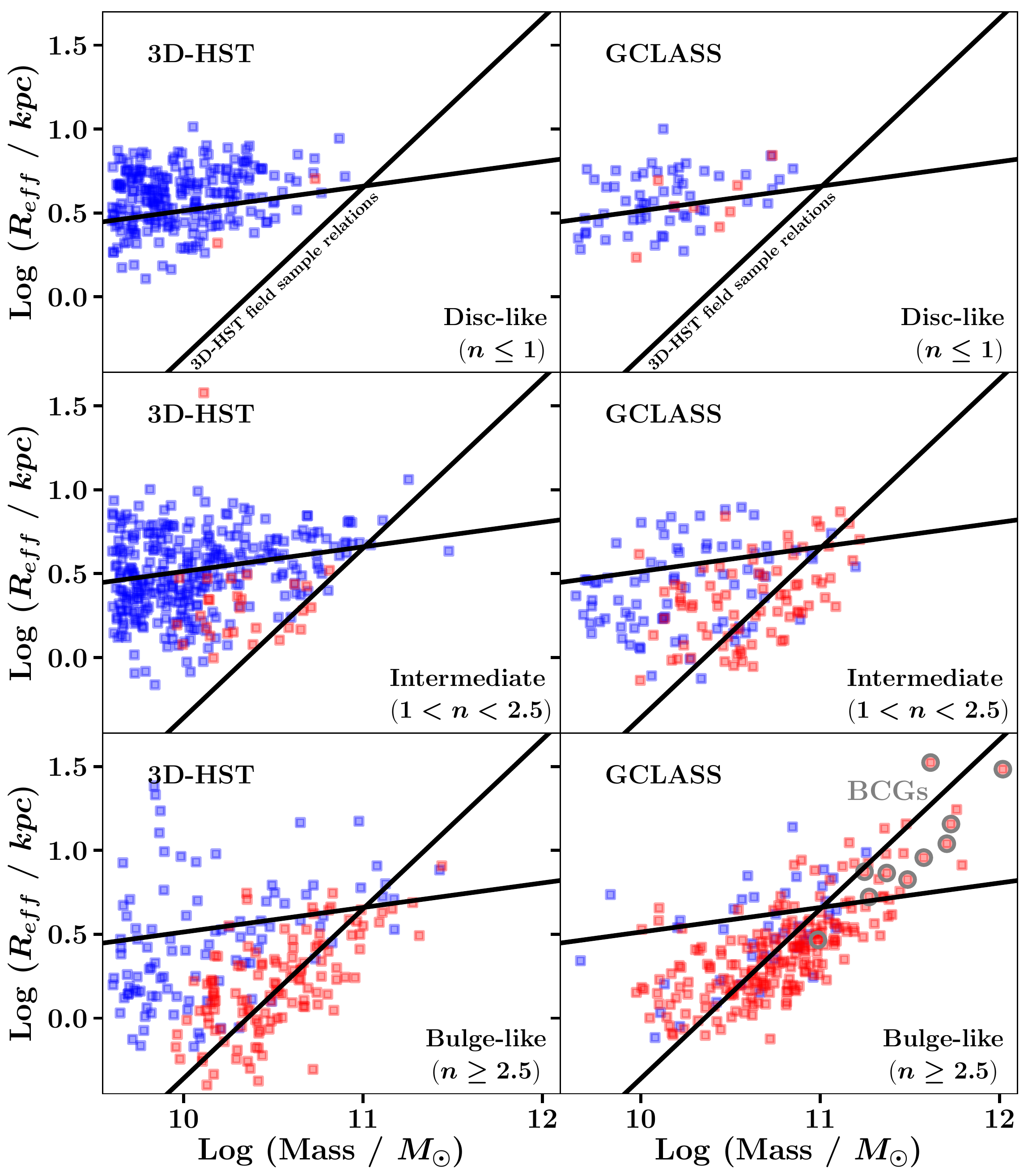}
    \caption{The stellar mass-size relation at $z~\mathtt{\sim}~1$ in both the field (left column) and cluster (right column) environment, split into three bins of S\'ersic index (top, middle and bottom rows) to track morphology. Blue squares indicate star-forming galaxies, red squares quiescent galaxies. Solid black lines indicate the field mass-size relations for {\it z} $\mathtt{\sim}$ 1 derived using results from \protect\cite{VanderWel2012,VanderWel2014}. Grey circles indicate the GCLASS BCGs.}
    \label{fig:morphology_mass_size_random}
\end{figure*}

\begin{figure}
	\includegraphics[width=\columnwidth]{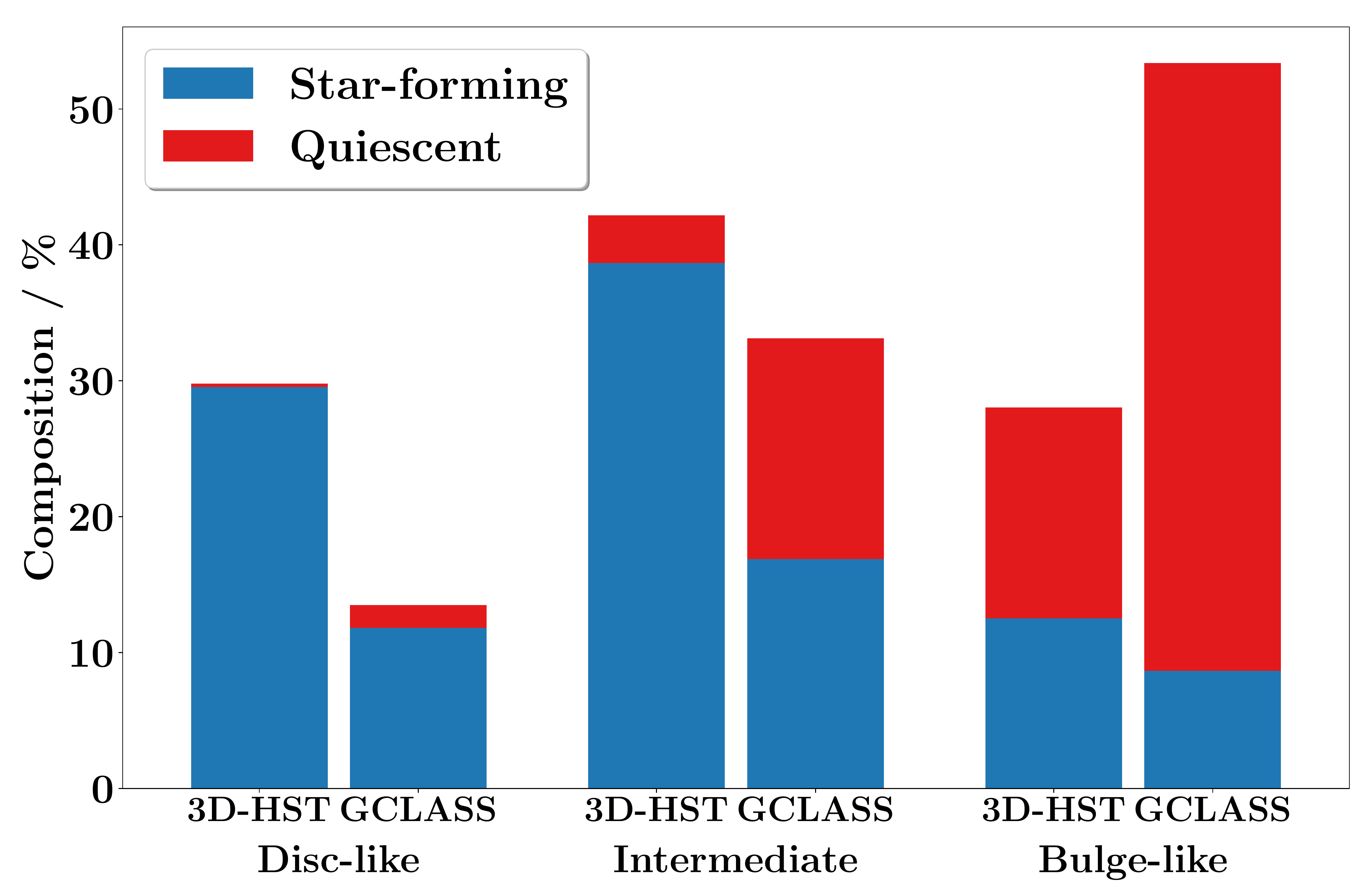}
    \caption{The morphological composition of both the field and cluster environments at fixed stellar mass. Quiescent intermediate-type and bulge-like galaxies are more popular in the cluster environment.}
    \label{fig:composition}
\end{figure}

We use the S\'ersic index measurements from the structural parameter estimation with GALFIT for GCLASS (see {Section~\ref{GALFIT_parameters}) and 3D-HST \citep{VanderWel2012} as a proxy to track morphology across the mass-size plane. Figure~\ref{fig:morphology_mass_size_random} shows the mass-size relation for both cluster and field samples in three bins of S\'ersic index. Disc-like and bulge-like morphologies tend to correlate with S\'ersic index (e.g. \citealt{Ravindranath2004}). We therefore use the S\'ersic index measurements as proxies, classing galaxies as ``disc-like" if they have $n\leqslant1$ and ``bulge-like" if they have $n\geqslant2.5$. We then categorise the rest of the galaxies as intermediate-types in a S\'ersic index bin between these two. To illustrate this comparison between the two environments, we truncate both samples to the mass completeness limits of GCLASS. The field sample used is one of the 1000 field samples that led to a cluster size offset close to the values found in Section~\ref{mass_size_sec}. These values are also stated in Figure~\ref{fig:offset_hists}.

Firstly, when comparing the location of galaxies on the mass-size plane based on their morphologies in both environments, it can be established that disc-like (bulge-like) galaxies lie closer to the star-forming (quiescent) field relation at $z~\mathtt{\sim}~1$. Alternatively, it shows that S\'ersic index is well-correlated with quiescence, as has been seen in other samples (e.g. \citealt{Franx2008,Bell2012}). Therefore, these field relations can be used as markers to track morphological changes across the mass-size plane. The mass-size relations for each morphological type are broadly the same in both environments. 

There is however a larger fraction of quiescent intermediates and quiescent bulge-like galaxies in the clusters compared to the field, whereas the disc-like population in both environments is dominated by star-forming galaxies. This is more clearly seen in Figure~\ref{fig:composition}. This excess population of quiescent intermediate-type galaxies points towards a cluster-specific process that quenches them more efficiently than if they resided in the field. A number of studies have identified an abundant population of ``passive spirals" or ``red discs" in clusters compared to the field \citep{VandenBergh1976,Goto2003,Moran2007,Gallazzi2008}. The most recent study of which is a study of the mass-size relation in the $z=0.44$ cluster MACS J1206.2-0847 \citep{Kuchner2017}. Here, they attribute its existence to cluster-specific quenching processes that lead to a fading of the stellar disc with respect to the inner bulge region. This would make the galaxy look smaller than a disc-like galaxy, which is probably why many of the quiescent intermediates lie in the region between the two field relations.

Furthermore, \cite{Kuchner2017} also found that this population was more prevalent in the regime \mbox{$R_{500}<R<R_{200}$}, which is where in-falling galaxies first feel the effects of the cluster's tidal field and start to be subjected to ``starvation" \citep{Moran2007}. This process can allow gas within the galaxy to escape \citep{Larson1980} as well as distort the distribution of gas. This was seen by \cite{Vogt2004}, who found that a larger population of asymmetric, stripped and quenched spirals are predominant in the hottest and richest of clusters based on H I properties. With the galaxy's gas supply removed, star formation is unable to continue in the galaxy. Eventually, the disc will fade further to the point where the bulge region of the galaxy will look brighter than the disc. The galaxy will then look bulge-like. Hence, this increased efficiency in quenching intermediate-types may be directly responsible for the larger fraction of quiescent bulges in the clusters. This result suggests that there is a direct morphological consequence of environmentally-driven quenching.

\section{Evolution of the cluster mass-size relation with redshift}
\label{BCG_growth}

\begin{figure*}
	\includegraphics[width=\textwidth]{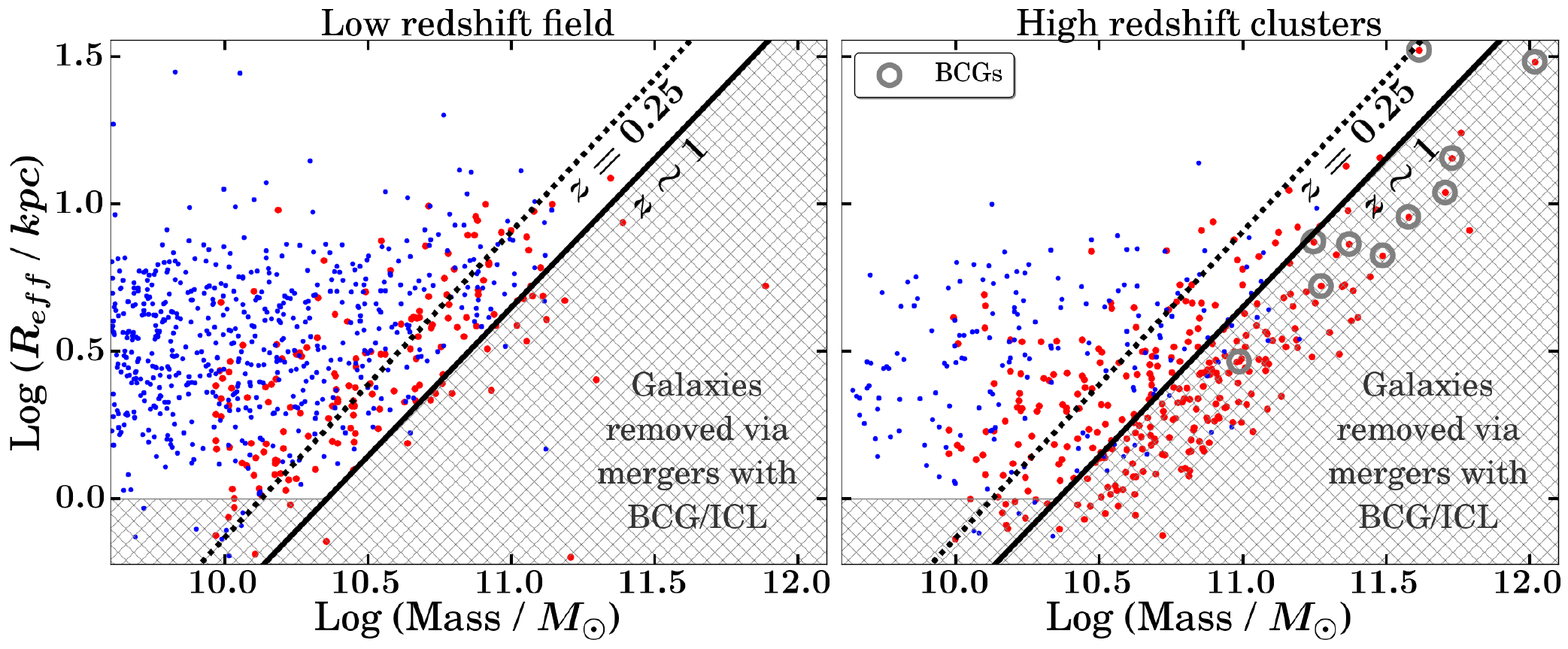}
    \caption{Left panel: $0<z<0.5$ field stellar mass-size relation from 3D-HST within the mass completeness limits of GCLASS. The black dashed line is the corresponding quiescent field relation calculated using results from \protect\cite{VanderWel2012} (see text). The solid black line shows the $z~\mathtt{\sim}~1$ quiescent field relation calculated in this study using 3D-HST measurements. Right panel: $z~\mathtt{\sim}~1$ cluster stellar mass-size relation from GCLASS within mass completeness limits. Lines are the same as in the left panel. Hatched region in both panels shows our toy model selection area for compact galaxies in GCLASS that should merge with their BCGs or become part of the ICL by low redshifts.}
    \label{fig:total_selection}
\end{figure*}

In Section~\ref{mass_size_sec}, we presented results that may support the hypothesis that minor mergers drive the size growth of galaxies in the field environment. However, the fact still remains that at low redshifts, small or negligible differences exist between the cluster and field stellar mass-size relations \citep{Weinmann2009,Maltby2010,Cebrian2014}. Since the intercept of the field mass-size relation increases with decreasing redshift \citep{VanderWel2014}, the same must therefore happen to the cluster mass-size relation for there to be such small differences between the cluster and field mass-size relations at low redshifts. This poses a problem for cluster galaxies. Whilst the observed decline in the number density of compact quiescent field galaxies with decreasing redshift (e.g \citealt{VanderWel2014}) could be explained by size growth via minor mergers, it cannot explain the increasing intercept of the cluster quiescent mass-size relation with decreasing redshift. This is because cluster galaxies are unlikely to grow in size via minor mergers. The size offsets towards smaller sizes measured in the GCLASS clusters support this hypothesis. Based on this hypothesis, as a galaxy cluster evolves, it will continually accrete new field galaxies, subsequently ``freezing" their size growth once they enter the cluster. Over time, larger field galaxies will be accreted, but the older, smaller field galaxies that were accreted at earlier times will still be present in the cluster. A build-up of compact cluster galaxies is created that will suppress increases in the intercept of the mass-size relation with decreasing redshift. To allow for the intercept of the cluster mass-size relation to significantly increase with decreasing redshift, there is only one option left. Since the compact cluster galaxies are unlikely to grow in size via minor mergers in the clusters, they need to be destroyed by some mechanism that is most likely cluster-specific.

\subsection{Toy model}

The two ways satellite galaxies in clusters can be destroyed are by either merging with their BCGs or being tidally disrupted into the ICL. We therefore construct a toy model to investigate 
whether the most compact cluster galaxies that exist in the GCLASS clusters at $z~\mathtt{\sim}~1$ can be destroyed by merging with their BCGs or being tidally disrupted into the ICL, such that the small differences observed between the cluster and field stellar mass-size relations at $z~\mathtt{\sim}~0$ can be achieved. The plausibility of this scenario depends upon several constraints.

\subsubsection{Constraints}
\label{constraints}
Recent studies have shown that BCGs increase their stellar mass by a factor of $\mathtt{\sim}2$ between $z~\mathtt{\sim}~1$ and $z~\mathtt{\sim}~0$ \citep{Lidman2012,Lidman2013,Lin2013, Bellstedt2016}. Therefore, the final stellar masses of the GCLASS BCGs at $z~\mathtt{\sim}~0$ cannot be more than approximately double their current stellar masses at $z~\mathtt{\sim}~1$ in the toy model. The remaining compact cluster galaxies left after this maximum stellar mass has been reached by the GCLASS BCGs would need to be tidally destroyed and contribute to the stellar mass of the ICL. However, there are also constraints on the stellar mass of the ICL at $z~\mathtt{\sim}~0$. Recent work at $z~\mathtt{\sim}~0.3-0.5$ has shown that the ICL stellar mass budget is $6-23\%$ of the entire cluster stellar mass contained within $\mathtt{\sim}R_{500}$ \citep{Presotto2014,Montes2014,Giallongo2014}. Hence if we assume the ICL stellar mass is negligible at $z~\mathtt{\sim}~1$, the total stellar mass of those compact cluster galaxies in the GCLASS clusters that need to be tidally destroyed can be no more than $6-23\%$ of their cluster's total stellar mass within $\mathtt{\sim}R_{500}$ at $z~\mathtt{\sim}~0$.

\subsubsection{Assumptions \& Sample selection}

We make the overall assumption that the majority of minor mergers a BCG will have between $z~\mathtt{\sim}~1$ and $z~\mathtt{\sim}~0$ will be with compact cluster galaxies that are already in place at $z~\mathtt{\sim}~1$. Simulations have also found that the majority of stars which end up in the BCG and ICL components fell into clusters before $z~\mathtt{\sim}~1$ \citep{Puchwein2010}. Galaxies which fell in later do not have enough time to sink towards the centre of the cluster and subsequently merge with the BCG. This is also a natural consequence of dynamical friction.

As mentioned earlier, low redshift work has shown that if there are any differences between the stellar mass-size relations in high and low density environments, they are very small. Therefore, we can assume that the stellar mass and size distribution of galaxies in both environments are similar. Hence we can use the low redshift field mass-size relation to identify a region which is required to be relatively absent of compact galaxies at $z~\mathtt{\sim}~0$. This region can then be used as our selection area for compact cluster galaxies in GCLASS that should be destroyed by $z~\mathtt{\sim}~0$. We show the low redshift field stellar mass-size relation from 3D-HST in the left-hand panel of Figure~\ref{fig:total_selection} accompanied with the low redshift quiescent field relation shown as the black dashed line\footnote{These measurements are selected in the same way as the field sample in this study, except a low redshift $UVJ$ colour sample selection is used to distinguish star-forming and quiescent galaxies.}. We also show the $z~\mathtt{\sim}~1$ quiescent field relation as the solid black line for comparison. The low redshift quiescent field relation was calculated using the same Bayesian technique used to calculate the $z~\mathtt{\sim}~1$ field relations in Section~\ref{mass_size_sec}. Due to the small sample of low redshift quiescent galaxies in 3D-HST with reliable size measurements (flag value 0 in the \cite{VanderWel2012} F160W catalogue, GALFIT measured $R_{eff}<50$~kpc and F160W magnitude~$<25$), we ran the fitting process 1000 times on the sample to capture the range of possible intercept and gradient values. The average line of all these fits is then used as the final quiescent field relation. Plotted points also include GALFIT fits that had a flag value of 1 in the \cite{VanderWel2012} F160W catalogue. The region marked with hatches signifies the region of the mass-size plane that is relatively absent of compact galaxies at $z~\mathtt{\sim}~0$. 

In the right panel of Figure~\ref{fig:total_selection}, we show this selection area and the low redshift field relation with respect to the GCLASS mass-size measurements. It can be seen that our chosen selection area is more populated in the $z~\mathtt{\sim}~1$ GCLASS clusters compared to the $z~\mathtt{\sim}~0$ field. For our toy model, we will select compact cluster galaxies from GCLASS that are within this region. We only apply our toy model to the sample of GCLASS galaxies that meet the mass completeness limits.

\subsubsection{Toy model results}

We use the analytic model for growth via minor mergers from \cite{Bezanson2009}, which dictates a growth in size of $R_{eff}\propto M_{*}^{2}$ after every minor merger event. To simplify our toy model, we do not use different growth prescriptions for minor mergers with star-forming/quiescent compact cluster galaxies to account for differing amounts of dissipation during the merger. Instead, we use a model appropriate for minor mergers between quiescent galaxies, since the majority of galaxies in the selection area are quiescent.

\begin{figure*}
	\includegraphics[width=\textwidth]{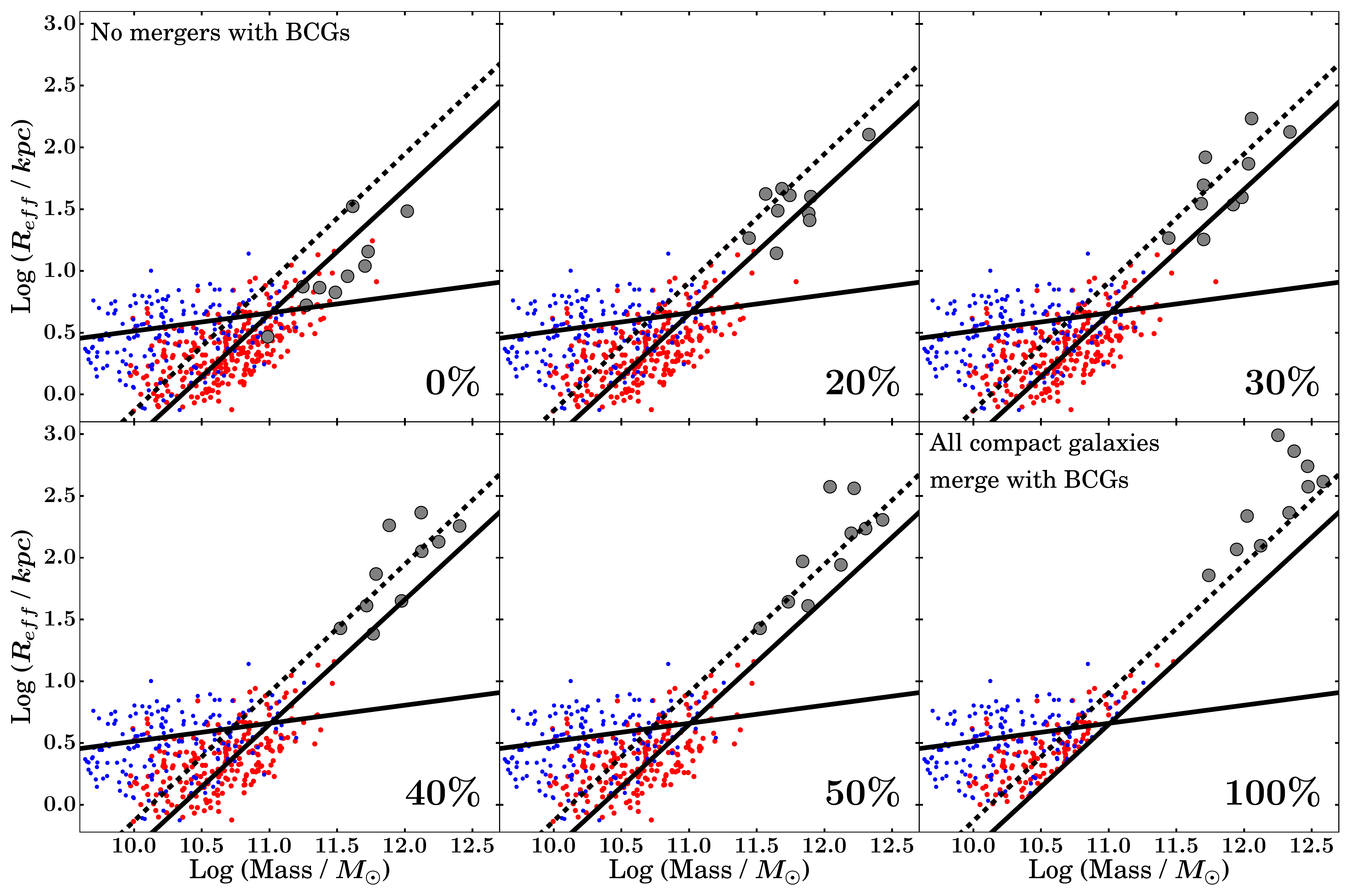}
    \caption{GCLASS BCG growth to low redshifts using the analytic model for growth via minor mergers from \protect\cite{Bezanson2009}. The first (top left) panel shows the original GCLASS stellar mass-size relation with the BCGs as large grey filled points. The last (bottom right) panel shows the position of the GCLASS BCGs in the extreme case where all compact cluster galaxies merge with their respective BCGs. In between, we show the best fit results for increasing percentage selections of compact cluster galaxies. Solid black lines are the $z~\mathtt{\sim}~1$ field relations calculated in this study and the black dashed line is the low redshift quiescent field relation calculated using results from \protect\cite{VanderWel2012} (see text).}
    \label{fig:BCG_model_FINAL}
\end{figure*}

In Figure~\ref{fig:BCG_model_FINAL}, we show the results for this growth prescription as well as the extreme case where all compact cluster galaxies merge with their respective BCGs. Compact galaxies that merge with their BCGs are randomly selected for each cluster from the hatched selection area (see right-hand panel of Figure~\ref{fig:total_selection}) for each percentage selection of compact galaxies shown in Figure~\ref{fig:BCG_model_FINAL}. The same random seed is used for each percentage, such that compact galaxies selected in smaller percentage prescriptions are part of the selection in higher percentage prescriptions plus additionally selected compact galaxies. This allows for a direct comparison between the different percentage prescriptions. It is clear that if all the compact cluster galaxies did merge with their respective BCGs ($100\%$ panel in Figure~\ref{fig:BCG_model_FINAL}), the sizes and stellar masses of the BCGs would be far too large compared to what has been observed in the local universe. 

We check what percentage of compact galaxies are required to merge with their BCG for each GCLASS cluster such that the BCG grows its stellar mass by a factor of $\mathtt{\sim}2$. This is the amount of stellar mass growth observed for BCGs between $z~\mathtt{\sim}~1$ and $z~\mathtt{\sim}~0$ for BCGs in clusters similar to those in GCLASS (see Section~\ref{constraints}). Once again, we randomly select compact galaxies from the hatched selection area shown in Figure~\ref{fig:total_selection} for each cluster to do this. Despite finding percentages ranging from $0\%$ (for SpARCS-1634) to $65\%$ (for SpARCS-1638), the average for the entire GCLASS sample is $40\%$. This suggests that no more than $40\%$ of compact galaxies on average in each GCLASS cluster can merge with their BCGs in our model. This fits nicely with the results of our toy model in Figure~\ref{fig:BCG_model_FINAL}, which suggests that most BCGs will follow the low redshift field quiescent stellar mass-size relation if $40\%$ of compact galaxies in each GCLASS cluster merge with their BCGs. Under our assumption that $100\%$ of compact cluster galaxies must be removed from the sample by $z~\mathtt{\sim}~0$, this implies the remaining $60\%$ of compact cluster galaxies are destroyed, most likely by tidal disruption into the ICL.

The HST imaging for all the clusters covers most, if not all of $R_{500}$ (see Section~\ref{hst_observations}). The total stellar mass of a GCLASS cluster is on average log($M_{*}/M_{\odot}$)$\mathtt{\sim}12.47$ assuming negligible ICL at $z~\mathtt{\sim}~1$. This is calculated by adding the stellar mass of all galaxies in each cluster above the mass-completeness limits and then finding the average of these 10 values. Based on the total stellar mass of the compact galaxies that did not merge with their BCGs in the $40\%$ prescription of our model, this destruction mechanism would lead to the build-up of an ICL by $z~\mathtt{\sim}~0$ with a stellar mass of log($M_{*}/M_{\odot}$)$\mathtt{\sim}11.96$. The average $M_{200}$ of the GCLASS sample is $\mathtt{\sim}~4.2\times10^{14}~M_{\odot}$ (see Table~\ref{tab:example_table}). A cluster of this mass at $z~\mathtt{\sim}~1$ is expected to grow by a factor of $\mathtt{\sim}4$ by $z~\mathtt{\sim}~0$ (see Figure 1 of \citealt{VanderBurg2014a}). If we assume the stellar mass contained within $\mathtt{\sim}R_{500}$ of the cluster increases by the same factor, this would lead to a total cluster stellar mass of log($M_{*}/M_{\odot}$)$\mathtt{\sim}13.07$ at $z~\mathtt{\sim}~0$. The ICL stellar mass fraction at $z~\mathtt{\sim}~0$ will therefore be $~\mathtt{\sim}8\%$ for a typical GCLASS cluster in our toy model. This agrees well with the $6-23\%$ found in more thorough studies of the ICL stellar mass fraction at low redshifts.

\subsubsection{Caveats}

Due to the simplicity of our toy model, there are other processes we have not considered that can alter the ICL stellar mass fraction. The first is that we only consider the possibility where compact cluster galaxies get completely destroyed into the ICL or completely merge with their BCGs. In a more realistic scenario, many compact cluster galaxies would be partially stripped into the ICL and partially merge with their BCGs. This partial stripping and partial merging can alter stellar mass fractions of the cluster and ICL. There is also the possibility that a small number of compact cluster galaxies survive between  $z~\mathtt{\sim}~1$ and $z~\mathtt{\sim}~0$, thereby reducing the ICL stellar mass fraction. Low redshift analogs of these compact cluster galaxies have been found in other works, suggesting that some do  survive (e.g. \citealt{Jiang2012}). It is also likely that some of the most massive quiescent cluster galaxies at $z~\mathtt{\sim}~1$ contribute stars to the ICL via tidal stripping of their outer regions as well \citep{Demaio2018}. This is an additional process that was not considered in our toy model which is capable of altering the ICL and total cluster stellar mass fractions. Our toy model is only applicable to cluster galaxies beyond the GCLASS mass completeness limits. Realistically, low mass galaxies below the mass completeness limits would also contribute stellar mass to the BCG/ICL components, but this contribution is likely to be sub-dominant. This is because most of the stellar mass in the GCLASS clusters is contained within galaxies that have a stellar mass of log($M_{*}/M_{\odot}$)$\mathtt{\sim}11$ (see Figure 2 in \citealt{VanderBurg2014}). 

We understand that realistically, BCGs will experience mergers with all types of galaxies - not just compact galaxies. We have therefore tested whether sampling from the entire cluster population rather than just the hatched region violates observational constraints of BCG stellar masses and the ICL stellar mass fraction at $z~\mathtt{\sim}~0$. This is done by re-running the toy model without confining the BCG/ICL selection area to the hatched region shown in Figure~\ref{fig:total_selection}. We find that a smaller percentage ($20\%$) of cluster galaxies are required to merge with their BCGs such that the BCGs grow their stellar mass by a factor of $\mathtt{\sim}2$, leading to a larger ICL stellar mass fraction of $18\%$ at $z~\mathtt{\sim}~0$. This ICL stellar mass fraction is still within the constraints found in low redshift studies (see Section~\ref{constraints}). The reason the hatched region was chosen was because the purpose of the toy model is to test whether there is a stellar mass budget available to account for the most compact galaxies that would place a drag on any increases in the intercept of the mass-size relation.

\subsubsection{Toy model conclusions}

The results of our toy model show that it is possible to achieve plausible ICL stellar mass fractions, BCG masses and BCG sizes by accounting for compact cluster galaxies in this way. This can consequently allow for the intercept of the cluster mass-size relation to increase with decreasing redshift, as has been observed. Recent work on the build-up of the ICL in clusters has shown that at $z<0.9$, the dominant route for ICL formation is the tidal stripping of galaxies with log($M_{*}/M_{\odot}$)>10.0. At least $75\%$ of the ICL luminosity in massive clusters is consistent in colour with originating from galaxies with log($M_{*}/M_{\odot}$)>10.4 \citep{Demaio2018}. It is therefore not surprising that we find the majority of compact galaxies -- which are all mostly quiescent -- in the GCLASS clusters with log($M_{*}/M_{\odot}$)>10.0 are required to contribute to the ICL to match low redshift observations of the stellar mass-size relation in our toy model. Cluster galaxies below these mass limits do not contribute as significantly to the BCG and ICL components, since most of the stellar mass in the GCLASS clusters is contained within galaxies that have a stellar mass of log($M_{*}/M_{\odot}$)$\mathtt{\sim}11$ (see Figure 2 in \citealt{VanderBurg2014}).

\subsection{Final remarks}

\begin{figure}
	\includegraphics[width=\columnwidth]{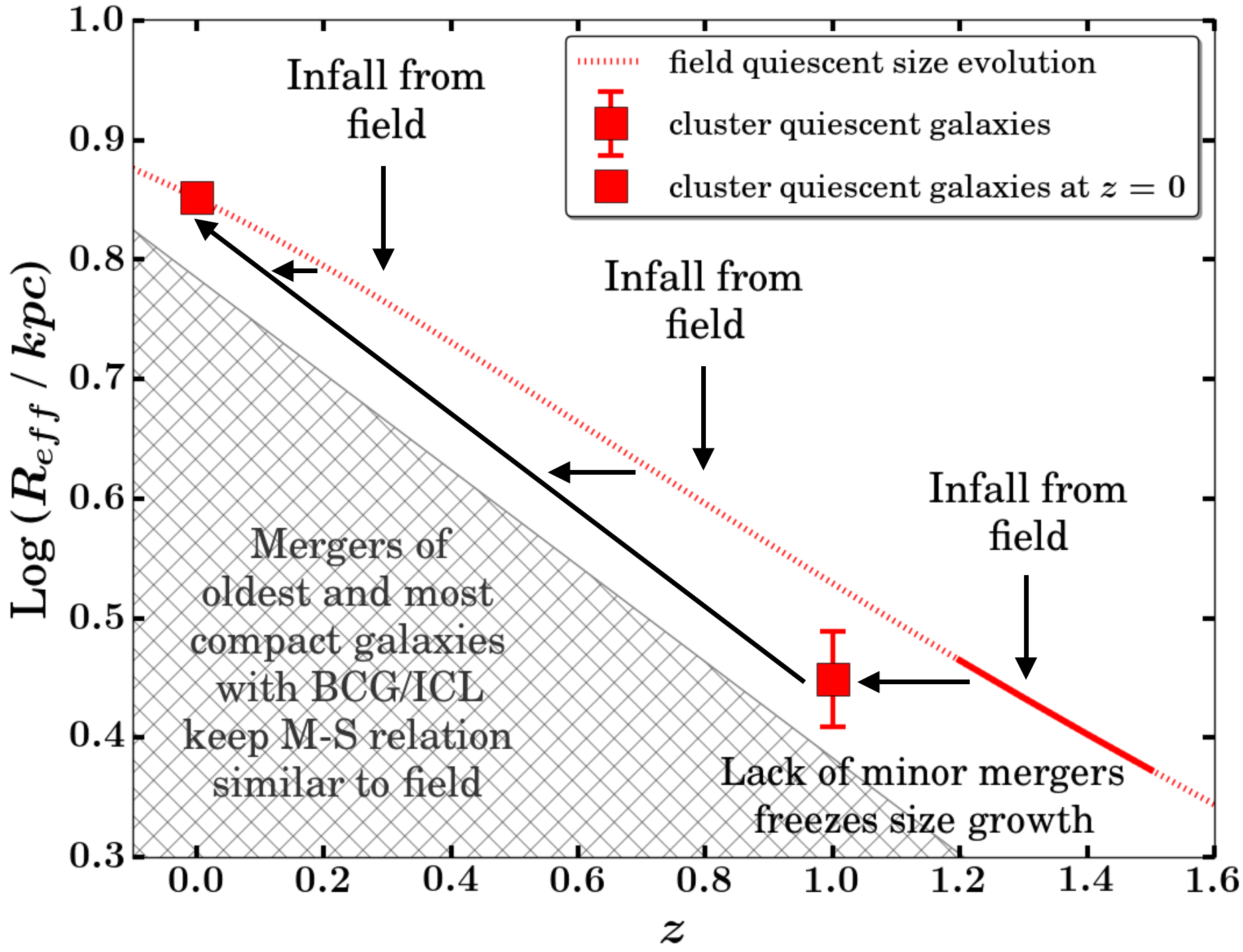}
    \caption{The physical processes occurring to reconcile the cluster quiescent stellar mass-size relation with the field quiescent stellar mass-size relation by $z~\mathtt{\sim}~0$. The field quiescent size evolution follows $R_{eff}~/$~kpc~$=~4.3~h(z)^{-1.29}$ (see Section~\ref{mass_size_sec}).}
    \label{fig:accretion}
\end{figure}

Figure~\ref{fig:accretion} summarises the physical processes thought to be occurring to reconcile the cluster quiescent stellar mass-size relation with the field quiescent stellar mass-size relation by low redshifts based on results found in this study. The preferred parameterisation for the evolution in the intercept of the quiescent mass-size relation with redshift for the 3D-HST fields is shown as the dotted red line (see Section~\ref{mass_size_sec}). The square point with the error bar shows the average size of a GCLASS quiescent cluster galaxy at $z~\mathtt{\sim}~1$. The current quiescent cluster galaxies we observe in the GCLASS clusters today likely fell into their clusters around $1.2\leqslant~z~\leqslant1.5$ (see Section~\ref{mass_size_sec}). At this time, they had average sizes that sit somewhere on the solid red line. Once these galaxies entered the cluster environment, they had their size growth suppressed due to the lack of minor mergers (based on the minor-mergers hypothesis). However, between the time of infall and observation of these galaxies, their field counterparts continued to grow via minor mergers in the field. Consequently, when the size offset between field and cluster is measured at $z~\mathtt{\sim}~1$, quiescent cluster galaxies are found to be smaller on average. At low redshifts however, negligible differences between cluster and field are seen. Between $z~\mathtt{\sim}~1$ and $z~\mathtt{\sim}~0$, most of the compact quiescent cluster galaxies are removed from the cluster sample via mergers with their BCGs or tidal destruction into the ICL. The sizes of these compact quiescent cluster galaxies would lie approximately in the hatched region shown in Figure~\ref{fig:accretion}, with some that are even smaller than Log~($R_{eff}/kpc)=0.3$. The removal of these small galaxies, and the continual addition of larger galaxies from the field, leads to smaller and smaller average size offsets between cluster and field quiescent galaxies with decreasing redshift. Furthermore, the minor mergers between the compact quiescent cluster galaxies and their BCGs allows the BCGs to grow disproportionately more in size, such that their sizes follow the low-redshift field quiescent mass-size relation by $z~\mathtt{\sim}~0$.
The combination of these physical processes lead to better agreement between the field and cluster quiescent stellar mass-size relations at low redshifts.

These results show that the observed increase in the intercept of the cluster mass-size relation can be explained by cluster-specific processes that we know to be occurring as a cluster evolves with time. Our results highlight the need for a careful comparison of the cluster mass-size relation over a broad redshift range, such that the physical mechanisms responsible for the evolution in the cluster mass-size relation can be directly observed, and therefore confirmed.

\section{Summary}
\label{conclusions}
Using the cluster environment as a laboratory, we tested whether minor mergers can explain the majority of the size growth observed in quiescent field galaxies.

To do this, we performed a comparison study of the stellar mass-size relation at $z~\mathtt{\sim}~1$ between cluster and field environments using the largest spectroscopically-confirmed sample of cluster galaxies at this redshift to date.

In a bid to reduce systematics as much as possible, observations, data reduction and stellar mass measurements were made in an almost identical fashion to those for our field sample from 3D-HST. A custom built GALFIT wrapper was developed in order to control for systematics between size and structural measurements for the comparative samples. The entire size determination method was first tested on a subset of the field sample and compared to existing published results to verify its reliability. Our method of size determination was proven to be highly reliable, with a systematic offset of $0.28\%$ (see Appendix~\ref{size_agreement}).

Our main conclusions are as follows:
\begin{enumerate}
\item Grism-derived redshifts for GCLASS have a precision of 2000 kms$^{-1}$, which is more than a factor of $4$ improvement over the photometric redshift precision. This allowed us to select a cluster membership sample from our grism data that was $\mathtt{\sim}90\%$ pure. Overall, this increased our cluster membership sample by $51\%$ of our spectroscopically confirmed sample with GMOS. This provided us with the largest sample of cluster galaxies at $z~\mathtt{\sim}~1$ for which stellar mass-size relation studies have been conducted.

\item Cluster galaxies are smaller than their field counterparts at fixed stellar mass. Average size offsets of \mbox{($-0.08\pm0.04$)~dex} and \mbox{($-0.07\pm0.01$)~dex} for quiescent and star-forming cluster galaxies are found respectively. The magnitude of these offsets are consistent with what is expected if minor mergers were the main drivers of galaxy size growth in the field.

\item There is a larger population of quiescent intermediate-type galaxies in the clusters compared to the field. These are likely to be galaxies undergoing environmental quenching -- most likely via ``starvation" -- such that their disc fades relative to their bulge over time. This is subsequently thought to be responsible for the larger population of quiescent bulge-like galaxies in the clusters compared to the field, suggesting a direct morphological consequence of environmental quenching. 

\item Using a toy model, we test whether the observed evolution in the intercept of the cluster mass-size relation with decreasing redshift can be explained by cluster-specific processes. We find that the small differences observed between the cluster and field stellar mass-size relations at low redshift can be achieved if $\mathtt{\sim}40\%$ of the compact cluster galaxies in GCLASS merge with their BCGs and the remaining $\mathtt{\sim}60\%$ become tidally disrupted into the ICL by $z~\mathtt{\sim}~0$. This leads to an ICL stellar mass fraction averaging $\mathtt{\sim}8\%$ at $z~\mathtt{\sim}~0$ for the GCLASS clusters. These results are consistent with the expected stellar mass growth of BCGs between $0\leqslant~z~\leqslant1$ and the expected stellar mass fraction of the ICL at $z~\mathtt{\sim}~0$.

\end{enumerate}

\section*{Acknowledgements}

J.M. is supported by the Science and Technology Facilities Council (STFC). This work is supported by the National Science Foundation through grant AST-1517863, by HST program numbers GO-13306, 13845, 13747, 13677/14327 and 15294,  and by grant number 80NSSC17K0019 issued through the NASA Astrophysics Data Analysis Program (ADAP). Support for program numbers GO-13306, 13845, 13747, 13677/14327 and 15294 was provided by NASA through a grant from the Space Telescope Science Institute, which is operated by the Association of Universities for Research in Astronomy, Incorporated, under NASA contract NAS5-26555. G.B. acknowledges funding of the Cosmic Dawn Center by the Danish National Research Foundation. R.D. gratefully acknowledges the support provided by the BASAL Center for Astrophysics and Associated Technologies (CATA) through grant AFB-170002. This research made use of \textsc{Astropy}, a community-developed core Python package for Astronomy \citep{TheAstropyCollaboration2018}. The python packages \textsc{Matplotlib} \citep{Hunter2007}, \textsc{Numpy}, \textsc{PyMC}, and \textsc{Scipy} were also extensively used.




\bibliographystyle{mnras}
\bibliography{Mendeley} 

\begin{thebibliography}{}
\makeatletter
\relax
\def\mn@urlcharsother{\let\do\@makeother \do\$\do\&\do\#\do\^\do\_\do\%\do\~}
\def\mn@doi{\begingroup\mn@urlcharsother \@ifnextchar [ {\mn@doi@}
  {\mn@doi@[]}}
\def\mn@doi@[#1]#2{\def\@tempa{#1}\ifx\@tempa\@empty \href
  {http://dx.doi.org/#2} {doi:#2}\else \href {http://dx.doi.org/#2} {#1}\fi
  \endgroup}
\def\mn@eprint#1#2{\mn@eprint@#1:#2::\@nil}
\def\mn@eprint@arXiv#1{\href {http://arxiv.org/abs/#1} {{\tt arXiv:#1}}}
\def\mn@eprint@dblp#1{\href {http://dblp.uni-trier.de/rec/bibtex/#1.xml}
  {dblp:#1}}
\def\mn@eprint@#1:#2:#3:#4\@nil{\def\@tempa {#1}\def\@tempb {#2}\def\@tempc
  {#3}\ifx \@tempc \@empty \let \@tempc \@tempb \let \@tempb \@tempa \fi \ifx
  \@tempb \@empty \def\@tempb {arXiv}\fi \@ifundefined
  {mn@eprint@\@tempb}{\@tempb:\@tempc}{\expandafter \expandafter \csname
  mn@eprint@\@tempb\endcsname \expandafter{\@tempc}}}

\bibitem[\protect\citeauthoryear{Allen et~al.,}{Allen et~al.}{2015}]{Allen2015}
Allen R.~J.,  et~al., 2015, \mn@doi [ApJ] {10.1088/0004-637X/806/1/3}, 806, 3

\bibitem[\protect\citeauthoryear{Andreon}{Andreon}{2018}]{Andreon2018}
Andreon S.,  2018, \mn@doi [A{\&}A] {10.1051/0004-6361/201832627}, 53

\bibitem[\protect\citeauthoryear{Barden, H{\"{a}}u{\ss}ler, Peng, McIntosh  \&
  Guo}{Barden et~al.}{2012}]{Barden2012}
Barden M.,  H{\"{a}}u{\ss}ler B.,  Peng C.~Y.,  McIntosh D.~H.,   Guo Y.,
  2012, \mn@doi [MNRAS] {10.1111/j.1365-2966.2012.20619.x}, 422, 449

\bibitem[\protect\citeauthoryear{Bassett et~al.,}{Bassett
  et~al.}{2013}]{Bassett2013}
Bassett R.,  et~al., 2013, \mn@doi [ApJ] {10.1088/0004-637X/770/1/58}, 770, 58

\bibitem[\protect\citeauthoryear{Bell et~al.,}{Bell et~al.}{2012}]{Bell2012}
Bell E.~F.,  et~al., 2012, \mn@doi [ApJ] {10.1088/0004-637X/753/2/167}, 753,
  167

\bibitem[\protect\citeauthoryear{Bellstedt et~al.,}{Bellstedt
  et~al.}{2016}]{Bellstedt2016}
Bellstedt S.,  et~al., 2016, \mn@doi [MNRAS] {10.1093/mnras/stw1184}, 460, 2862

\bibitem[\protect\citeauthoryear{Bertin \& Arnouts}{Bertin \&
  Arnouts}{1996}]{Bertin1996}
Bertin E.,  Arnouts S.,  1996, \mn@doi [A{\&}AS] {10.1051/aas:1996164}, 117,
  393

\bibitem[\protect\citeauthoryear{Bezanson, van Dokkum, Tal, Marchesini, Kriek,
  Franx  \& Coppi}{Bezanson et~al.}{2009}]{Bezanson2009}
Bezanson R.,  van Dokkum P.~G.,  Tal T.,  Marchesini D.,  Kriek M.,  Franx M.,
   Coppi P.,  2009, ApJ, 697, 1290

\bibitem[\protect\citeauthoryear{Bezanson et~al.,}{Bezanson
  et~al.}{2016}]{Bezanson2016}
Bezanson R.,  et~al., 2016, \mn@doi [ApJ] {10.3847/0004-637X/822/1/30}, 822, 30

\bibitem[\protect\citeauthoryear{Biviano, van~der Burg, Muzzin, Sartoris,
  Wilson  \& Yee}{Biviano et~al.}{2016}]{Biviano2016}
Biviano A.,  van~der Burg R. F.~J.,  Muzzin A.,  Sartoris B.,  Wilson G.,   Yee
  H. K.~C.,  2016, \mn@doi [A{\&}A] {10.1051/0004-6361/201628697}, 594, 1

\bibitem[\protect\citeauthoryear{Bluck, Conselice, Buitrago, Gr{\"{u}}tzbauch,
  Hoyos, Mortlock  \& Bauer}{Bluck et~al.}{2012}]{Bluck2012}
Bluck A.~F.,  Conselice C.~J.,  Buitrago F.,  Gr{\"{u}}tzbauch R.,  Hoyos C.,
  Mortlock A.,   Bauer A.~E.,  2012, AJ, 747

\bibitem[\protect\citeauthoryear{Brammer, Dokkum  \& Coppi}{Brammer
  et~al.}{2008}]{Brammer2008}
Brammer G.~B.,  Dokkum P. G.~V.,   Coppi P.,  2008, ApJ, 686, 1503

\bibitem[\protect\citeauthoryear{Brammer et~al.,}{Brammer
  et~al.}{2012}]{Brammer2012}
Brammer G.~B.,  et~al., 2012, ApJS, 200, 13

\bibitem[\protect\citeauthoryear{Calzetti, Armus, Bohlin, Kinney, Koornneef  \&
  Storchi-Bergmann}{Calzetti et~al.}{2000}]{Calzetti1999}
Calzetti D.,  Armus L.,  Bohlin R.~C.,  Kinney A.~L.,  Koornneef J.,
  Storchi-Bergmann T.,  2000, \mn@doi [ApJ] {10.1086/308692}, 533, 682

\bibitem[\protect\citeauthoryear{Cappellari}{Cappellari}{2013}]{Cappellari2013}
Cappellari M.,  2013, \mn@doi [ApJL] {10.1088/2041-8205/778/1/L2}, 778

\bibitem[\protect\citeauthoryear{Carollo et~al.,}{Carollo
  et~al.}{2013}]{Carollo2013}
Carollo C.~M.,  et~al., 2013, \mn@doi [ApJ] {10.1088/0004-637X/773/2/112}, 773

\bibitem[\protect\citeauthoryear{Cebri{\'{a}}n \& Trujillo}{Cebri{\'{a}}n \&
  Trujillo}{2014}]{Cebrian2014}
Cebri{\'{a}}n M.,  Trujillo I.,  2014, \mn@doi [MNRAS] {10.1093/mnras/stu1375},
  444, 682

\bibitem[\protect\citeauthoryear{Chabrier}{Chabrier}{2003}]{Chabrier2003}
Chabrier G.,  2003, \mn@doi [PASP] {10.1086/376392}, 115, 763

\bibitem[\protect\citeauthoryear{Chan et~al.,}{Chan et~al.}{2018}]{Chan2018}
Chan J. C.~C.,  et~al., 2018, \mn@doi [arXiv:1802.01609] {arXiv:1802.01609}

\bibitem[\protect\citeauthoryear{Cooper et~al.,}{Cooper
  et~al.}{2012}]{Cooper2012}
Cooper M.~C.,  et~al., 2012, \mn@doi [MNRAS]
  {10.1111/j.1365-2966.2011.19938.x}, 419, 3018

\bibitem[\protect\citeauthoryear{DeMaio, Gonzalez, Zabludoff, Zaritsky, Connor,
  Donahue  \& Mulchaey}{DeMaio et~al.}{2018}]{Demaio2018}
DeMaio T.,  Gonzalez A.~H.,  Zabludoff A.,  Zaritsky D.,  Connor T.,  Donahue
  M.,   Mulchaey J.~S.,  2018, \mn@doi [MNRAS] {10.1093/mnras/stx2946}, 474,
  3009

\bibitem[\protect\citeauthoryear{Delahaye et~al.,}{Delahaye
  et~al.}{2017}]{Delahaye2017}
Delahaye A.~G.,  et~al., 2017, ApJ, 843, 126

\bibitem[\protect\citeauthoryear{Delaye et~al.,}{Delaye
  et~al.}{2014}]{Delaye2014a}
Delaye L.,  et~al., 2014, \mn@doi [MNRAS] {10.1093/mnras/stu496}, 441, 203

\bibitem[\protect\citeauthoryear{Demarco et~al.,}{Demarco
  et~al.}{2010}]{Demarco2010}
Demarco R.,  et~al., 2010, \mn@doi [ApJ] {10.1088/0004-637X/711/2/1185}, 711,
  1185

\bibitem[\protect\citeauthoryear{Fakhouri, Ma  \& Boylan-Kolchin}{Fakhouri
  et~al.}{2010}]{Fakhouri2010}
Fakhouri O.,  Ma C.~P.,   Boylan-Kolchin M.,  2010, \mn@doi [MNRAS]
  {10.1111/j.1365-2966.2010.16859.x}, 406, 2267

\bibitem[\protect\citeauthoryear{Fan, Lapi, De~Zotti  \& Danese}{Fan
  et~al.}{2008}]{Fan2008}
Fan L.,  Lapi A.,  De~Zotti G.,   Danese L.,  2008, \mn@doi [ApJ]
  {10.1086/595784}, 689, L101

\bibitem[\protect\citeauthoryear{Fern{\'{a}}ndez-Lorenzo, Sulentic,
  Verdes-Montenegro  \& Argudo-Fernandez}{Fern{\'{a}}ndez-Lorenzo
  et~al.}{2013}]{Lorenzo2013}
Fern{\'{a}}ndez-Lorenzo M.,  Sulentic J.,  Verdes-Montenegro L.,
  Argudo-Fernandez M.,  2013, \mn@doi [MNRAS] {10.1093/mnras/stt1020}, 434, 325

\bibitem[\protect\citeauthoryear{Ferreras et~al.,}{Ferreras
  et~al.}{2014}]{Ferreras2014}
Ferreras I.,  et~al., 2014, \mn@doi [MNRAS] {10.1093/mnras/stu1425}, 444, 906

\bibitem[\protect\citeauthoryear{Franx, Van~Dokkum, Schreiber, Wuyts  \&
  Labb{\'{e}}}{Franx et~al.}{2008}]{Franx2008}
Franx M.,  Van~Dokkum P.~G.,  Schreiber N. M.~F.,  Wuyts S.,   Labb{\'{e}} I.,
  2008, ApJ, 688

\bibitem[\protect\citeauthoryear{Galametz et~al.,}{Galametz
  et~al.}{2016}]{Galametz2016}
Galametz A.,  et~al., 2016, \mn@doi [ApJS] {10.3847/1538-4365/228/1/7}, 206, 19

\bibitem[\protect\citeauthoryear{Gallazzi et~al.,}{Gallazzi
  et~al.}{2008}]{Gallazzi2008}
Gallazzi A.,  et~al., 2008, \mn@doi [ApJ] {10.1088/0004-637X/690/2/1883}, 690,
  1883

\bibitem[\protect\citeauthoryear{Gardner et~al.,}{Gardner
  et~al.}{2006}]{Gardner2006}
Gardner J.~P.,  et~al., 2006, \mn@doi [Space Science Reviews]
  {10.1007/s11214-006-8315-7}, 123, 485

\bibitem[\protect\citeauthoryear{Giallongo et~al.,}{Giallongo
  et~al.}{2014}]{Giallongo2014}
Giallongo E.,  et~al., 2014, \mn@doi [ApJ] {10.1088/0004-637X/781/1/24}, 781,
  24

\bibitem[\protect\citeauthoryear{Gladders \& Yee}{Gladders \&
  Yee}{2000}]{Gladders2000}
Gladders M.~D.,  Yee H. K.~C.,  2000, \mn@doi [AJ] {10.1086/301557}, 120, 2148

\bibitem[\protect\citeauthoryear{Goto et~al.,}{Goto et~al.}{2003}]{Goto2003}
Goto T.,  et~al., 2003, \mn@doi [PASJ] {10.1093/pasj/55.4.757}, 55, 31

\bibitem[\protect\citeauthoryear{Green et~al.,}{Green et~al.}{2012}]{Green2012}
Green J.,  et~al., 2012, \mn@doi [arXiv:1208.4012] {arXiv:1208.4012}, pp
  2182--2184

\bibitem[\protect\citeauthoryear{Grogin et~al.,}{Grogin
  et~al.}{2011}]{Grogin2011}
Grogin N.~A.,  et~al., 2011, ApJS, 197, 35

\bibitem[\protect\citeauthoryear{Hilz, Naab, Ostriker, Thomas, Burkert  \&
  Jesseit}{Hilz et~al.}{2012}]{Hilz2012}
Hilz M.,  Naab T.,  Ostriker J.~P.,  Thomas J.,  Burkert A.,   Jesseit R.,
  2012, MNRAS, 3136, 3119

\bibitem[\protect\citeauthoryear{Hopkins, Bundy, Murray, Quataert, Lauer  \&
  Ma}{Hopkins et~al.}{2009a}]{Hopkins2009b}
Hopkins P.~F.,  Bundy K.,  Murray N.,  Quataert E.,  Lauer T.~R.,   Ma C.~P.,
  2009a, MNRAS, 398, 898

\bibitem[\protect\citeauthoryear{Hopkins, Hernquist, Cox, Keres  \&
  Wuyts}{Hopkins et~al.}{2009b}]{Hopkins2009c}
Hopkins P.~F.,  Hernquist L.,  Cox T.~J.,  Keres D.,   Wuyts S.,  2009b,
  \mn@doi [ApJ] {10.1088/0004-637X/691/2/1424}, 691, 1424

\bibitem[\protect\citeauthoryear{Huertas-Company et~al.,}{Huertas-Company
  et~al.}{2013a}]{Huertas-Company2013}
Huertas-Company M.,  et~al., 2013a, \mn@doi [MNRAS] {10.1093/mnras/sts150},
  428, 1715

\bibitem[\protect\citeauthoryear{Huertas-Company, Shankar, Mei, Bernardi,
  Aguerri, Meert  \& Vikram}{Huertas-Company
  et~al.}{2013b}]{Huertas-Company2013a}
Huertas-Company M.,  Shankar F.,  Mei S.,  Bernardi M.,  Aguerri J. A.~L.,
  Meert A.,   Vikram V.,  2013b, \mn@doi [ApJ] {10.1088/0004-637X/779/1/29},
  779, 29

\bibitem[\protect\citeauthoryear{Hunter}{Hunter}{2007}]{Hunter2007}
Hunter J.~D.,  2007, Computing in Science and Engineering, 9, 90

\bibitem[\protect\citeauthoryear{Jiang, Van~Dokkum, Bezanson  \& Franx}{Jiang
  et~al.}{2012}]{Jiang2012}
Jiang F.,  Van~Dokkum P.,  Bezanson R.,   Franx M.,  2012, \mn@doi [ApJL]
  {10.1088/2041-8205/749/1/L10}, 749

\bibitem[\protect\citeauthoryear{Kelkar, Arag{\'{o}}n-Salamanca, Gray, Maltby,
  Vulcani, De~Lucia, Poggianti  \& Zaritsky}{Kelkar et~al.}{2015}]{Kelkar2015}
Kelkar K.,  Arag{\'{o}}n-Salamanca A.,  Gray M.~E.,  Maltby D.,  Vulcani B.,
  De~Lucia G.,  Poggianti B.~M.,   Zaritsky D.,  2015, \mn@doi [MNRAS]
  {10.1093/mnras/stv670}, 450, 1246

\bibitem[\protect\citeauthoryear{Kelly}{Kelly}{2007}]{Kelly2007}
Kelly B.~C.,  2007, ApJ, 665, 1489

\bibitem[\protect\citeauthoryear{Koekemoer et~al.,}{Koekemoer
  et~al.}{2011}]{Koekemoer2011}
Koekemoer A.~M.,  et~al., 2011, ApJS, 197, 36

\bibitem[\protect\citeauthoryear{Kriek, van Dokkum, Labbe, Franx, Illingworth,
  Marchesini  \& Quadri}{Kriek et~al.}{2009}]{Kriek2009}
Kriek M.,  van Dokkum P.~G.,  Labbe I.,  Franx M.,  Illingworth G.~D.,
  Marchesini D.,   Quadri R.~F.,  2009, \mn@doi [ApJ]
  {10.1088/0004-637X/700/1/221}, 700, 221

\bibitem[\protect\citeauthoryear{Kuchner, Ziegler, Verdugo, Bamford  \&
  H{\"{a}}u{\ss}ler}{Kuchner et~al.}{2017}]{Kuchner2017}
Kuchner U.,  Ziegler B.,  Verdugo M.,  Bamford S.,   H{\"{a}}u{\ss}ler B.,
  2017, \mn@doi [A{\&}A] {10.1051/0004-6361/201630252}, 54, 1

\bibitem[\protect\citeauthoryear{Lani et~al.,}{Lani et~al.}{2013}]{Lani2013}
Lani C.,  et~al., 2013, \mn@doi [MNRAS] {10.1093/mnras/stt1275}, 435, 207

\bibitem[\protect\citeauthoryear{Larson, Tinsley  \& Caldwell}{Larson
  et~al.}{1980}]{Larson1980}
Larson R.~B.,  Tinsley B.~M.,   Caldwell C.~N.,  1980, \mn@doi [ApJ]
  {10.1086/157917}, 237, 692

\bibitem[\protect\citeauthoryear{Laureijs et~al.,}{Laureijs
  et~al.}{2011}]{Report2011}
Laureijs R.,  et~al., 2011, \mn@doi [arXiv:1110.3193] {arXiv:1110.3193}

\bibitem[\protect\citeauthoryear{Lidman et~al.,}{Lidman
  et~al.}{2012}]{Lidman2012}
Lidman C.,  et~al., 2012, \mn@doi [MNRAS] {10.1111/j.1365-2966.2012.21984.x},
  427, 550

\bibitem[\protect\citeauthoryear{Lidman et~al.,}{Lidman
  et~al.}{2013}]{Lidman2013}
Lidman C.,  et~al., 2013, \mn@doi [MNRAS] {10.1093/mnras/stt777}, 433, 825

\bibitem[\protect\citeauthoryear{Lin, Brodwin, Gonzalez, Bode, Eisenhardt,
  Stanford  \& Vikhlinin}{Lin et~al.}{2013}]{Lin2013}
Lin Y.~T.,  Brodwin M.,  Gonzalez A.~H.,  Bode P.,  Eisenhardt P.~R.,  Stanford
  S.~A.,   Vikhlinin A.,  2013, \mn@doi [ApJ] {10.1088/0004-637X/771/1/61}, 771

\bibitem[\protect\citeauthoryear{L{\'{o}}pez-Sanjuan, Balcells,
  P{\'{e}}rez-Gonz{\'{a}}lez, Barro, Garc{\'{i}}a-Dab{\'{o}}, Gallego  \&
  Zamorano}{L{\'{o}}pez-Sanjuan et~al.}{2009}]{Lopez-Sanjuan2009}
L{\'{o}}pez-Sanjuan C.,  Balcells M.,  P{\'{e}}rez-Gonz{\'{a}}lez P.~G.,  Barro
  G.,  Garc{\'{i}}a-Dab{\'{o}} C.~E.,  Gallego J.,   Zamorano J.,  2009,
  \mn@doi [A{\&}A] {10.1051/0004-6361/200911923}, 501, 505

\bibitem[\protect\citeauthoryear{Maltby et~al.,}{Maltby
  et~al.}{2010}]{Maltby2010}
Maltby D.~T.,  et~al., 2010, \mn@doi [MNRAS]
  {10.1111/j.1365-2966.2009.15953.x}, 402, 282

\bibitem[\protect\citeauthoryear{Merritt}{Merritt}{1985}]{Merritt1985}
Merritt D.,  1985, ApJ, 289, 18

\bibitem[\protect\citeauthoryear{Momcheva et~al.,}{Momcheva
  et~al.}{2016}]{Momcheva2016}
Momcheva I.~G.,  et~al., 2016, ApJS, 225

\bibitem[\protect\citeauthoryear{Montes \& Trujillo}{Montes \&
  Trujillo}{2014}]{Montes2014}
Montes M.,  Trujillo I.,  2014, \mn@doi [ApJ] {10.1088/0004-637X/794/2/137},
  794

\bibitem[\protect\citeauthoryear{Moore, Katz, Lake, Dressler  \& Oemler}{Moore
  et~al.}{1996}]{Moore1996}
Moore B.,  Katz N.,  Lake G.,  Dressler A.,   Oemler A.,  1996, \mn@doi
  [Nature] {10.1038/379613a0}, 379, 613

\bibitem[\protect\citeauthoryear{Moore, Lake  \& Katz}{Moore
  et~al.}{1998}]{Moore1998}
Moore B.,  Lake G.,   Katz N.,  1998, \mn@doi [ApJ] {10.1086/305264}, 495, 139

\bibitem[\protect\citeauthoryear{Moran, Ellis, Treu, Smith, Rich  \&
  Smail}{Moran et~al.}{2007}]{Moran2007}
Moran S.~M.,  Ellis R.~S.,  Treu T.,  Smith G.~P.,  Rich R.~M.,   Smail I.,
  2007, \mn@doi [ApJ] {10.1086/522303}, 671, 1503

\bibitem[\protect\citeauthoryear{Morishita et~al.,}{Morishita
  et~al.}{2017}]{Morishita2016}
Morishita T.,  et~al., 2017, \mn@doi [ApJ] {10.3847/1538-4357/835/2/254}, 835,
  254

\bibitem[\protect\citeauthoryear{Muzzin, Wilson, Lacy, Yee  \& Stanford}{Muzzin
  et~al.}{2008}]{Muzzin2008}
Muzzin A.,  Wilson G.,  Lacy M.,  Yee H. K.~C.,   Stanford S.~A.,  2008,
  \mn@doi [ApJ] {10.1086/591542}, 686, 966

\bibitem[\protect\citeauthoryear{Muzzin et~al.,}{Muzzin
  et~al.}{2009}]{Muzzin2009}
Muzzin A.,  et~al., 2009, \mn@doi [ApJ] {10.1088/0004-637X/698/2/1934}, 698,
  1934

\bibitem[\protect\citeauthoryear{Muzzin et~al.,}{Muzzin
  et~al.}{2012}]{Muzzin2012}
Muzzin A.,  et~al., 2012, ApJ, 746, 188

\bibitem[\protect\citeauthoryear{Naab, Johansson  \& Ostriker}{Naab
  et~al.}{2009}]{Naab2009}
Naab T.,  Johansson P.~H.,   Ostriker J.~P.,  2009, ApJ, 699, L178

\bibitem[\protect\citeauthoryear{Newman, Ellis, Andreon, Treu, Raichoor  \&
  Trinchieri}{Newman et~al.}{2014}]{Newman2014b}
Newman A.~B.,  Ellis R.~S.,  Andreon S.,  Treu T.,  Raichoor A.,   Trinchieri
  G.,  2014, \mn@doi [ApJ] {10.1088/0004-637X/788/1/51}, 788, 51

\bibitem[\protect\citeauthoryear{Oser, Naab, Ostriker  \& Johansson}{Oser
  et~al.}{2012}]{Oser2012}
Oser L.,  Naab T.,  Ostriker J.~P.,   Johansson P.~H.,  2012, ApJ, 744, 63

\bibitem[\protect\citeauthoryear{Papovich et~al.,}{Papovich
  et~al.}{2012a}]{Papovich2012}
Papovich C.,  et~al., 2012a, ] {10.1088/0004-637X/750/2/93}, 750, 93

\bibitem[\protect\citeauthoryear{Papovich et~al.,}{Papovich
  et~al.}{2012b}]{Papovich2012a}
Papovich C.,  et~al., 2012b, \mn@doi [ApJ] {10.1088/0004-637X/750/2/93}, 750,
  93

\bibitem[\protect\citeauthoryear{Patel, Holden, Kelson, Franx, van~der Wel  \&
  Illingworth}{Patel et~al.}{2012}]{Patel2012}
Patel S.~G.,  Holden B.~P.,  Kelson D.~D.,  Franx M.,  van~der Wel A.,
  Illingworth G.~D.,  2012, \mn@doi [ApJL] {10.1088/2041-8205/748/2/L27}, 748,
  L27

\bibitem[\protect\citeauthoryear{Peng, Ho, Impey  \& Rix}{Peng
  et~al.}{2002}]{Peng2002b}
Peng C.~Y.,  Ho L.~C.,  Impey C.~D.,   Rix H.-W.,  2002, \mn@doi [AJ]
  {10.1086/340952}, 124, 266

\bibitem[\protect\citeauthoryear{Peng, Ho, Impey  \& Rix}{Peng
  et~al.}{2010}]{Peng2010a}
Peng C.~Y.,  Ho L.~C.,  Impey C.~D.,   Rix H.-W.,  2010, \mn@doi [AJ]
  {10.1088/0004-6256/139/6/2097}, 139, 2097

\bibitem[\protect\citeauthoryear{Planck Collaboration~XIII}{Planck
  Collaboration~XIII}{2016}]{Planck2015}
Planck Collaboration~XIII P.,  2016, \mn@doi [A{\&}A]
  {10.1051/0004-6361/201525830}, 13

\bibitem[\protect\citeauthoryear{Poggianti et~al.,}{Poggianti
  et~al.}{2013}]{Poggianti2013}
Poggianti B.~M.,  et~al., 2013, \mn@doi [ApJ] {10.1088/0004-637X/762/2/77},
  762, 77

\bibitem[\protect\citeauthoryear{Presotto et~al.,}{Presotto
  et~al.}{2014}]{Presotto2014}
Presotto V.,  et~al., 2014, \mn@doi [A{\&}A] {10.1051/0004-6361/201323251}, 565

\bibitem[\protect\citeauthoryear{Puchwein, Springel, Sijacki  \&
  Dolag}{Puchwein et~al.}{2010}]{Puchwein2010}
Puchwein E.,  Springel V.,  Sijacki D.,   Dolag K.,  2010, \mn@doi [MNRAS]
  {10.1111/j.1365-2966.2010.16786.x}, 406, 936

\bibitem[\protect\citeauthoryear{Raichoor et~al.,}{Raichoor
  et~al.}{2012}]{Raichoor2012}
Raichoor A.,  et~al., 2012, \mn@doi [ApJ] {10.1088/0004-637X/745/2/130}, 745,
  130

\bibitem[\protect\citeauthoryear{Ravindranath et~al.,}{Ravindranath
  et~al.}{2004}]{Ravindranath2004}
Ravindranath S.,  et~al., 2004, \mn@doi [ApJ] {10.1086/382952}, 604, L9

\bibitem[\protect\citeauthoryear{Rettura et~al.,}{Rettura
  et~al.}{2010}]{Rettura2010}
Rettura A.,  et~al., 2010, \mn@doi [ApJ] {10.1088/0004-637X/709/1/512}, 709,
  512

\bibitem[\protect\citeauthoryear{Rubin et~al.,}{Rubin et~al.}{2017}]{Rubin2017}
Rubin D.,  et~al., 2017, \mn@doi [arXiv:1707.04606v2] {arXiv:1707.04606v2}

\bibitem[\protect\citeauthoryear{Saracco, Gargiulo, Ciocca  \&
  Marchesini}{Saracco et~al.}{2017}]{Saracco2017}
Saracco P.,  Gargiulo A.,  Ciocca F.,   Marchesini D.,  2017, \mn@doi [A{\&}A]
  {10.1051/0004-6361/201628866}, 597, A122

\bibitem[\protect\citeauthoryear{Shen, Mo, White, Blanton, Kauffmann, Voges,
  Brinkmann  \& Csabai}{Shen et~al.}{2003}]{Shen2003}
Shen S.,  Mo H.~J.,  White S. D.~M.,  Blanton M.~R.,  Kauffmann G.,  Voges W.,
  Brinkmann J.,   Csabai I.,  2003, \mn@doi [MNRAS]
  {10.1046/j.1365-8711.2003.06740.x}, 343, 978

\bibitem[\protect\citeauthoryear{Skelton et~al.,}{Skelton
  et~al.}{2014}]{Skelton2014a}
Skelton R.~E.,  et~al., 2014, \mn@doi [ApJS] {10.1088/0067-0049/214/2/24}, 214,
  24

\bibitem[\protect\citeauthoryear{Strazzullo et~al.,}{Strazzullo
  et~al.}{2013}]{Strazzullo2013}
Strazzullo V.,  et~al., 2013, \mn@doi [ApJ] {10.1088/0004-637X/772/2/118}, 772

\bibitem[\protect\citeauthoryear{Sweet et~al.,}{Sweet et~al.}{2017}]{Sweet2016}
Sweet S.~M.,  et~al., 2017, \mn@doi [MNRAS] {10.1093/mnras/stw2411}, 464, 24

\bibitem[\protect\citeauthoryear{{The Astropy Collaboration}}{{The Astropy
  Collaboration}}{2018}]{TheAstropyCollaboration2018}
{The Astropy Collaboration} 2018, \mn@doi [arXiv:1801.02634v2]
  {arXiv:1801.02634v2}

\bibitem[\protect\citeauthoryear{Trujillo, Ferreras  \& de~la Rosa}{Trujillo
  et~al.}{2011}]{Trujillo2011}
Trujillo I.,  Ferreras I.,   de~la Rosa I.~G.,  2011, MNRAS, 415, 3903

\bibitem[\protect\citeauthoryear{Valentinuzzi et~al.,}{Valentinuzzi
  et~al.}{2010a}]{Valentinuzzi2010a}
Valentinuzzi T.,  et~al., 2010a, \mn@doi [ApJ] {10.1088/0004-637X/712/1/226},
  712, 226

\bibitem[\protect\citeauthoryear{Valentinuzzi et~al.,}{Valentinuzzi
  et~al.}{2010b}]{Valentinuzzi2010b}
Valentinuzzi T.,  et~al., 2010b, \mn@doi [ApJ] {10.1088/2041-8205/721/1/L19},
  721, L19

\bibitem[\protect\citeauthoryear{Vogt, Haynes, Giovanelli  \& Herter}{Vogt
  et~al.}{2004}]{Vogt2004}
Vogt N.~P.,  Haynes M.~P.,  Giovanelli R.,   Herter T.,  2004, \mn@doi [AJ]
  {10.1086/420702}, 127, 3300

\bibitem[\protect\citeauthoryear{Weinmann, Kauffmann, Van Den~Bosch, Pasquali,
  McIntosh, Mo, Yang  \& Guo}{Weinmann et~al.}{2009}]{Weinmann2009}
Weinmann S.~M.,  Kauffmann G.,  Van Den~Bosch F.~C.,  Pasquali A.,  McIntosh
  D.~H.,  Mo H.,  Yang X.,   Guo Y.,  2009, \mn@doi [MNRAS]
  {10.1111/j.1365-2966.2009.14412.x}, 394, 1213

\bibitem[\protect\citeauthoryear{Williams, Quadri, Franx, van Dokkum  \&
  Labbe}{Williams et~al.}{2009}]{Williams2008}
Williams R.~J.,  Quadri R.~F.,  Franx M.,  van Dokkum P.,   Labbe I.,  2009,
  \mn@doi [ApJ] {10.1088/0004-637X/691/2/1879}, 691, 1879

\bibitem[\protect\citeauthoryear{Wilson et~al.,}{Wilson
  et~al.}{2009}]{Wilson2009a}
Wilson G.,  et~al., 2009, \mn@doi [ApJ] {10.1088/0004-637X/698/2/1943}, 698,
  1943

\bibitem[\protect\citeauthoryear{Wuyts et~al.,}{Wuyts et~al.}{2007}]{Wuyts2007}
Wuyts S.,  et~al., 2007, \mn@doi [ApJ] {10.1086/509708}, 655, 51

\bibitem[\protect\citeauthoryear{Yoon, Im  \& Kim}{Yoon
  et~al.}{2017}]{Yoon2017}
Yoon Y.,  Im M.,   Kim J.-W.,  2017, \mn@doi [ApJ]
  {10.3847/1538-4357/834/1/73}, 834, 73

\bibitem[\protect\citeauthoryear{van Dokkum et~al.,}{van Dokkum
  et~al.}{2010}]{VanDokkum2010}
van Dokkum P.~G.,  et~al., 2010, \mn@doi [ApJ] {10.1088/0004-637X/709/2/1018},
  709, 1018

\bibitem[\protect\citeauthoryear{van~den Bergh}{van~den
  Bergh}{1976}]{VandenBergh1976}
van~den Bergh S.,  1976, \mn@doi [ApJ] {10.1086/154452}, 206, 883

\bibitem[\protect\citeauthoryear{van~der Burg, Muzzin, Hoekstra, Lidman,
  Rettura, Wilson  \& Yee}{van~der Burg et~al.}{2013}]{VanderBurg2013}
van~der Burg R. F.~J.,  Muzzin A.,  Hoekstra H.,  Lidman C.,  Rettura A.,
  Wilson G.,   Yee H. K.~C.,  2013, A{\&}A, 557, A15

\bibitem[\protect\citeauthoryear{van~der Burg, Muzzin, Hoekstra, Wilson, Lidman
   \& Yee}{van~der Burg et~al.}{2014}]{VanderBurg2014}
van~der Burg R.~F.~J.,  Muzzin A.,  Hoekstra H.,  Wilson G.,  Lidman C.,   Yee
  H.~K.~C.,  2014, \mn@doi [A{\&}A] {10.1051/0004-6361/201322771}, 561, A79

\bibitem[\protect\citeauthoryear{van~der Burg, Hoekstra, Muzzin, Sif{\'{o}}n,
  Balogh  \& McGee}{van~der Burg et~al.}{2015}]{VanderBurg2014a}
van~der Burg R. F.~J.,  Hoekstra H.,  Muzzin A.,  Sif{\'{o}}n C.,  Balogh
  M.~L.,   McGee S.~L.,  2015, \mn@doi [A{\&}A] {10.1051/0004-6361/201425460},
  577

\bibitem[\protect\citeauthoryear{van~der Wel, Bell, Van Den~Bosch, Gallazzi  \&
  Rix}{van~der Wel et~al.}{2009}]{VanDerWel2009}
van~der Wel A.,  Bell E.~F.,  Van Den~Bosch F.~C.,  Gallazzi A.,   Rix H.~W.,
  2009, \mn@doi [ApJ] {10.1088/0004-637X/698/2/1232}, 698, 1232

\bibitem[\protect\citeauthoryear{van~der Wel et~al.,}{van~der Wel
  et~al.}{2012}]{VanderWel2012}
van~der Wel A.,  et~al., 2012, \mn@doi [ApJS] {10.1088/0067-0049/203/2/24},
  203, 24

\bibitem[\protect\citeauthoryear{van~der Wel et~al.,}{van~der Wel
  et~al.}{2014}]{VanderWel2014}
van~der Wel A.,  et~al., 2014, \mn@doi [ApJ] {10.1088/0004-637X/788/1/28}, 788,
  28

\makeatother
\end{thebibliography}



\appendix

\section{A summary of previous results on the cluster vs. field stellar mass-size relation for quiescent galaxies}
\label{previous}

In Table~\ref{tab:previous_tab} and Figure~\ref{fig:previous_work}, we summarise results from studies since 2009 of the cluster versus field stellar mass-size relation for quiescent galaxies.

\begin{table*}
\centering
	\caption{A summary of results from cluster versus field stellar mass-size relation studies of quiescent galaxies since 2009. Studies are listed by redshift in ascending order.}
	\label{tab:previous_tab}
	\begin{tabular}{lccccc}
    	\hline
        Reference & Redshift & Stellar Mass & Offset & Error & Method of measurement\\
        & & [log($M_{*}/M_{\odot}$)] & [w.r.t the field]\\
        \hline
        \cite{Lorenzo2013} & $0$ & $9.0\lesssim$~log($M_{*}$)~$\lesssim11.5$ & 0 & - & zeropoint\\
        \cite{Cappellari2013} & 0.0231 & $10.5\lesssim$~log($M_{*}$)~$\lesssim11.3$ & 0 & $\pm4\%$ & size distributions$^{a}$ \\
                \cite{Huertas-Company2013a} & $0<z<0.09$ & $10.5\lesssim$~log($M_{*}$)~$\lesssim11.8$ & 0 & $30-40\%$ & median relations\\
                        \cite{Valentinuzzi2010a} & $0.04<z<0.07$ & $10.5\lesssim$~log($M_{*}$)~$\lesssim11.6$ & $-0.1$~dex & - & median sizes\\
        \cite{Cebrian2014} & $0<z<0.12$ & $9.0\lesssim$~log($M_{*}$)~$\lesssim11.0$ & $-4.0\%$ & $\pm0.8\%$ & mean sizes\\
        \cite{Poggianti2013} & $0.03<z<0.11$ & $10.5\lesssim$~log($M_{*}$)~$\lesssim11.7$ & $-1\sigma$ & - & median relations\\
        \cite{Weinmann2009} & $0.01<z<0.2$ & $9.8\lesssim$~log($M_{*}$)~$\lesssim12.0$ & 0 & - & median sizes\\
        \cite{Yoon2017} & $0.1<z<0.15$ & $11.2\lesssim$~log($M_{*}$)~$\lesssim11.8$ & $20-40\%$ & - & median sizes\\
        \cite{Maltby2010} & $0.167$ & $9.0\lesssim$~log($M_{*}$)~$\lesssim11.0$ & 0 & - & mean sizes\\
        \cite{Kuchner2017} & $0.44$ & $9.8\lesssim$~log($M_{*}$)~$\lesssim10.8$ & $2-5\sigma$ & - & mean sizes\\
        \cite{Morishita2016} & $0.2<z<0.7$ & $7.8\lesssim$~log($M_{*}$)~$\lesssim12.0$ & $-7\%$ & $\pm3\%$ & multidimensional analysis$^{b}$\\
        \cite{Huertas-Company2013} & $0.2<z<1$ & $10.7\lesssim$~log($M_{*}$)~$\lesssim11.8$ & 0 & - & mass-normalised radii$^{c}$\\
        \cite{Kelkar2015} & $0.4<z<0.8$ & $10.2\lesssim$~log($M_{*}$)~$\lesssim12.0$ & 0 & $10-20\%$ & K-S test$^{d}$\\
        \cite{Cooper2012} & $0.4<z<1.2$ & $10.0\lesssim$~log($M_{*}$)~$\lesssim11.0$ & $0.54$~kpc  & $\pm0.22$~kpc & H-L estimator$^{e}$\\
        Matharu et al. (2019) & $0.86<z<1.34$ & $10.0\lesssim$~log($M_{*}$)~$\lesssim12.0$ & $-0.08$~dex & $\pm0.04$~dex & mean sizes\\
        \cite{Sweet2016} & $1.067$ & $9.5\lesssim$~log($M_{*}$)~$\lesssim12.0$ & 0 & - & fitted relations$^{f}$\\
        \cite{Delaye2014a} & $0.8<z<1.5$ & $10.5\lesssim$~log($M_{*}$)~$\lesssim11.5$ & $30-40\%$ & - & mass-normalised radii$^{g}$\\
        \cite{Rettura2010} & $1.237$ & $10.7\lesssim$~log($M_{*}$)~$\lesssim11.2$ & 0 & $<20\%$ & fitted relations\\
        \cite{Raichoor2012} & $1.3$ & $11.0\lesssim$~log($M_{*}$)~$\lesssim11.5$ & $-~30-50\%$ & - & size ratio distributions$^{h}$\\
        \cite{Saracco2017} & $1.3$ & $9.0\lesssim$~log($M_{*}$)~$\lesssim12.0$ & 0 & - & K-S test\\
        \cite{Chan2018} & $1.39<z<1.61$ & $10.5\lesssim$~log($M_{*}$)~$\lesssim11.8$ & $24\%$ & - & median sizes\\
        \cite{Lani2013} & $1<z<2$ & $11.3\lesssim$~log($M_{*}$)~$\lesssim11.8$ & $48\%$ & $\pm25$ & mean sizes\\
        \cite{Bassett2013} & $1.6$ & $10.3\lesssim$~log($M_{*}$)~$\lesssim11.3$ & $0.6$~kpc & $2\sigma$ & median sizes\\
        \cite{Papovich2012} & $1.62$ & $10.5\lesssim$~log($M_{*}$)~$\lesssim11.0$ & $0.7$~kpc & $+1.7, -0.9^{i}$~kpc & median sizes\\
        \cite{Newman2014b} & $1.8$ & $10.8\lesssim$~log($M_{*}$)~$\lesssim11.7$ & $0.01$~dex & $0.09$~dex & mean sizes\\
        \cite{Strazzullo2013} & $2$ & $10.0\lesssim$~log($M_{*}$)~$\lesssim11.5$ & $2\times$~larger & $\pm0.08^{j}$ & mean sizes$^{k}$\\
        \cite{Andreon2018} & $2$ & $10.7\lesssim$~log($M_{*}$)~$\lesssim11.5$ & $3\times$ larger & - & mass-normalised sizes$^{l}$\\
        \cite{Allen2015} & $2.1$ & $10.7\lesssim$~log($M_{*}$)~$\lesssim11.5$ & $0.36$~kpc & $0.69$~kpc$^{}$ & mass-normalised radii\\
	\end{tabular}
    \begin{flushleft}
    $^{a}$ difference in the mean of the size probability distributions.\\
    $^{b}$ A relation where size is a function of multiple variables -- not just stellar mass -- is defined, leaving environment as the only distinguishing factor.\\
    $^{c}$ $r_{e} \times (10^{11} M_{\odot}/M_{*})$\\
    $^{d}$ Kolmogorov-Smirnov tests were carried out to estimate the probability, p, that the field and cluster samples were derived from the same size distribution. Environmental differences were considered significant if p < 0.05 ($>2\sigma$ significance).\\
    $^{e}$ Hodge-Lehmann estimator of the mean. This is given by the median value of the mean computed over all pairs of galaxies in the sample.\\
    $^{f}$ Comparison of fitted mass-size relation to that of \cite{VanderWel2014} in their Figure 9.\\
    $^{g}$ $r_{e} / M_{11}^{0.57}$, where $M_{11}=M_{*}/10^{11}M_{\odot}$\\
    $^{h}$ Peak of the size ratio normalised distributions, where the size ratio is $r_{e}/r_{e, Valen}$. $r_{e, Valen}$ is the half-light radius as predicted by the \cite{Valentinuzzi2010a} mass-size relation. \\
    $^{i}$ Error propagation of the interquartile ranges stated for field and cluster medians in this work.\\
    $^{j}$ Error propagation of the mean $r_{e}/r_{e, Shen2003}$ errors.
    $^{k}$ difference in the average size ratio $r_{e}/r_{e, Shen2003}$, where $r_{e, Shen2003}$ is the average half-light radius as given by the \cite{Shen2003} mass-size relation.\\
    $^{l}$ Comparison of the mean galaxy size at log$(M_{*}/M_{\odot})=11$.\\
    \end{flushleft}
\end{table*}

\begin{figure*}
	\includegraphics[width=\textwidth]{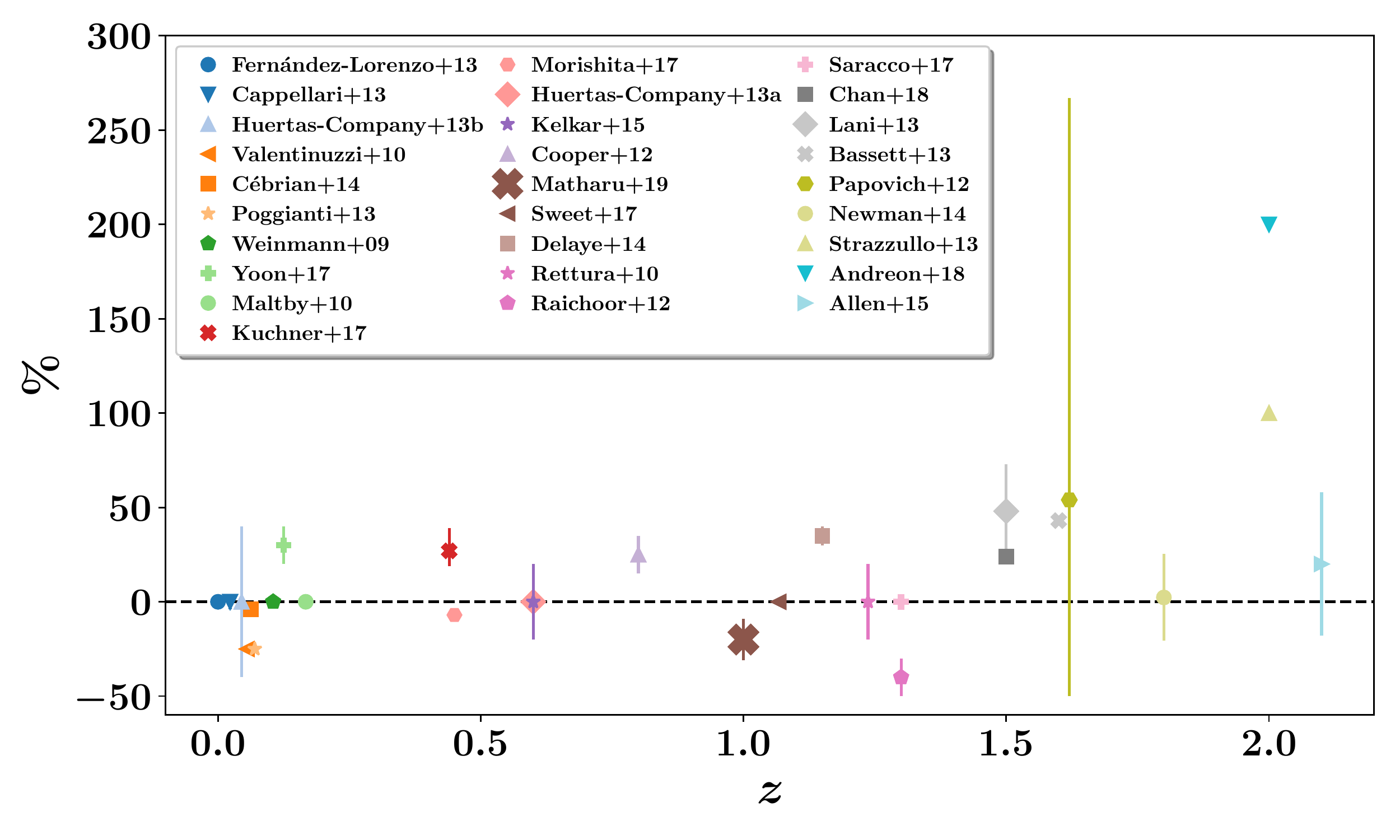}
    \caption{A summary of results from cluster versus field stellar mass-size relation studies of quiescent galaxies since 2009. Positive percentages indicate larger sizes in clusters with respect to the field. Error bars are shown for those works which stated errors that could easily be converted into percentages. Legend labels are arranged by redshift in ascending order. See Table~\ref{tab:previous_tab} for more details.}
    \label{fig:previous_work}
\end{figure*}

\section{SExtractor settings for GCLASS F140W mosaics}
\label{GCLASS_sex}
\begin{table}
	\centering
	\caption{SExtractor configuration file parameters for the GCLASS F140W mosaics}
	\label{tab:GCLASS_sex_config}
	\begin{tabular}{lc}
		\hline
		Parameter & Value\\
		\hline
		$\mathtt{CATALOG}$\verb!_!$\mathtt{TYPE}$ &  $\mathtt{ASCII}$\verb!_!$\mathtt{HEAD}$\\
		$\mathtt{DETECT}$\verb!_!$\mathtt{TYPE}$ & $\mathtt{CCD}$\\
		$\mathtt{DETECT}$\verb!_!$\mathtt{MINAREA}$ & $\mathtt{5}$\\
                $\mathtt{DETECT}$\verb!_!$\mathtt{THRESH}$ & $\mathtt{2.0}$\\
                $\mathtt{ANALYSIS}$\verb!_!$\mathtt{THRESH}$ & $\mathtt{2.0}$\\
                $\mathtt{FILTER}$ & $\mathtt{Y}$\\
                $\mathtt{FILTER}$\verb!_!$\mathtt{NAME}$ & $\mathtt{gauss}$\verb!_!$\mathtt{1.5}$\verb!_!$\mathtt{3x3.conv}$\\
                $\mathtt{DEBLEND}$\verb!_!$\mathtt{NTHRESH}$ & $\mathtt{32}$\\
                $\mathtt{DEBLEND}$\verb!_!$\mathtt{MINCONT}$ & $\mathtt{0.005}$\\
                $\mathtt{CLEAN}$ & $\mathtt{Y}$\\
                $\mathtt{CLEAN}$\verb!_!$\mathtt{PARAM}$ & $\mathtt{1.0}$\\
                $\mathtt{MASK}$\verb!_!$\mathtt{TYPE}$ & $\mathtt{CORRECT}$\\
                $\mathtt{PHOT}$\verb!_!$\mathtt{APERTURES}$ & $\mathtt{5}$\\
                $\mathtt{PHOT}$\verb!_!$\mathtt{AUTOPARAMS}$ & $\mathtt{2.5, 3.5}$\\
                $\mathtt{PHOT}$\verb!_!$\mathtt{PETROPARAMS}$ & $\mathtt{2.0, 3.5}$\\
             $\mathtt{PHOT}$\verb!_!$\mathtt{FLUXFRAC}$ & $\mathtt{0.5}$\\
                $\mathtt{SATUR}$\verb!_!$\mathtt{LEVEL}$ & $\mathtt{50000.0}$\\
                $\mathtt{SATUR}$\verb!_!$\mathtt{KEY}$ & $\mathtt{SATURATE}$\\
                $\mathtt{MAG}$\verb!_!$\mathtt{ZEROPOINT}$ & $\mathtt{26.45}$\\
                $\mathtt{MAG}$\verb!_!$\mathtt{GAMMA}$ & $\mathtt{4.0}$\\
                $\mathtt{GAIN}$ & $\mathtt{0.0}$\\
                $\mathtt{GAIN}$\verb!_!$\mathtt{KEY}$ & $\mathtt{GAIN}$\\
                $\mathtt{PIXEL}$\verb!_!$\mathtt{SCALE}$ & $\mathtt{0.06}$\\
                $\mathtt{SEEING}$\verb!_!$\mathtt{FWHM}$ & $\mathtt{0.23}$\\
                $\mathtt{STARNNW}$\verb!_!$\mathtt{NAME}$ & $\mathtt{default.nnw}$\\
                $\mathtt{BACK}$\verb!_!$\mathtt{SIZE}$ & $\mathtt{64}$\\
                $\mathtt{BACK}$\verb!_!$\mathtt{FILTERSIZE}$ & $\mathtt{3}$\\
                $\mathtt{BACKPHOTO}$\verb!_!$\mathtt{TYPE}$ & $\mathtt{LOCAL}$\\
               $\mathtt{WEIGHT}$\verb!_!$\mathtt{TYPE}$ & $\mathtt{MAP}$\verb!_!$\mathtt{WEIGHT}$\\
                $\mathtt{WEIGHT}$\verb!_!$\mathtt{GAIN}$ & $\mathtt{Y}$\\ 	
                $\mathtt{MEMORY}$\verb!_!$\mathtt{OBJSTACK}$ & $\mathtt{3000}$\\
                $\mathtt{MEMORY}$\verb!_!$\mathtt{PIXSTACK}$ & $\mathtt{300000}$\\
                $\mathtt{MEMORY}$\verb!_!$\mathtt{BUFSIZE}$ & $\mathtt{1024}$\\
		\hline
	\end{tabular}
\end{table}
Table ~\ref{tab:GCLASS_sex_config} lists the values for the parameters in the SExtractor configuration file that was run on the GCLASS F140W mosaics.

\section{Size comparison test with van der Wel et al. (2012)}
\label{size_agreement}

We took the F160W mosaic for CANDELS-COSMOS, and measured the half-light radii of all the galaxies in the field using our method of size determination (see Section ~\ref{size_determination}). The reason we did not use the F140W mosaic -  which would have made the comparison to GCLASS more direct - was because the published structural parameters for all the CANDELS fields in 3D-HST were measured using the F125W and F160W mosaics, sometimes accompanied with measurements in F098m or F105W \citep{VanderWel2012}. Both F125W and F160W span approximately half of the wavelength range covered by F140W. In Appendix~\ref{filter_difference}, we explore how the differing filters for field and cluster samples affect results in this paper. In Figure~\ref{fig:size_size}, we plot our half-light radii results against those from \cite{VanderWel2012} in F160W for the same set of galaxies with \mbox{$0.86<z<1.34$}, \mbox{F160W magnitude~$<25$},  GALFIT measured \mbox{$R_{eff}<50$~kpc}, flag value of 0 (see Section~\ref{sample_selection} for the reasoning behind this selection) and stellar masses within the mass completeness limits of our study (log$(M_{*}/M_{\odot})>9.96$ and log$(M_{*}/M_{\odot})>9.60$ for quiescent and star-forming galaxies, respectively).

Despite the differing methods of size determination, there are no systematics present. The mean offset between the two sets of measurements is $0.28\%$. Divergence from agreement at larger half-light radii is due to poor signal-to-noise ratio (F160W magnitude $\mathtt{\sim}25$).

\begin{figure}
	\includegraphics[width=\columnwidth]{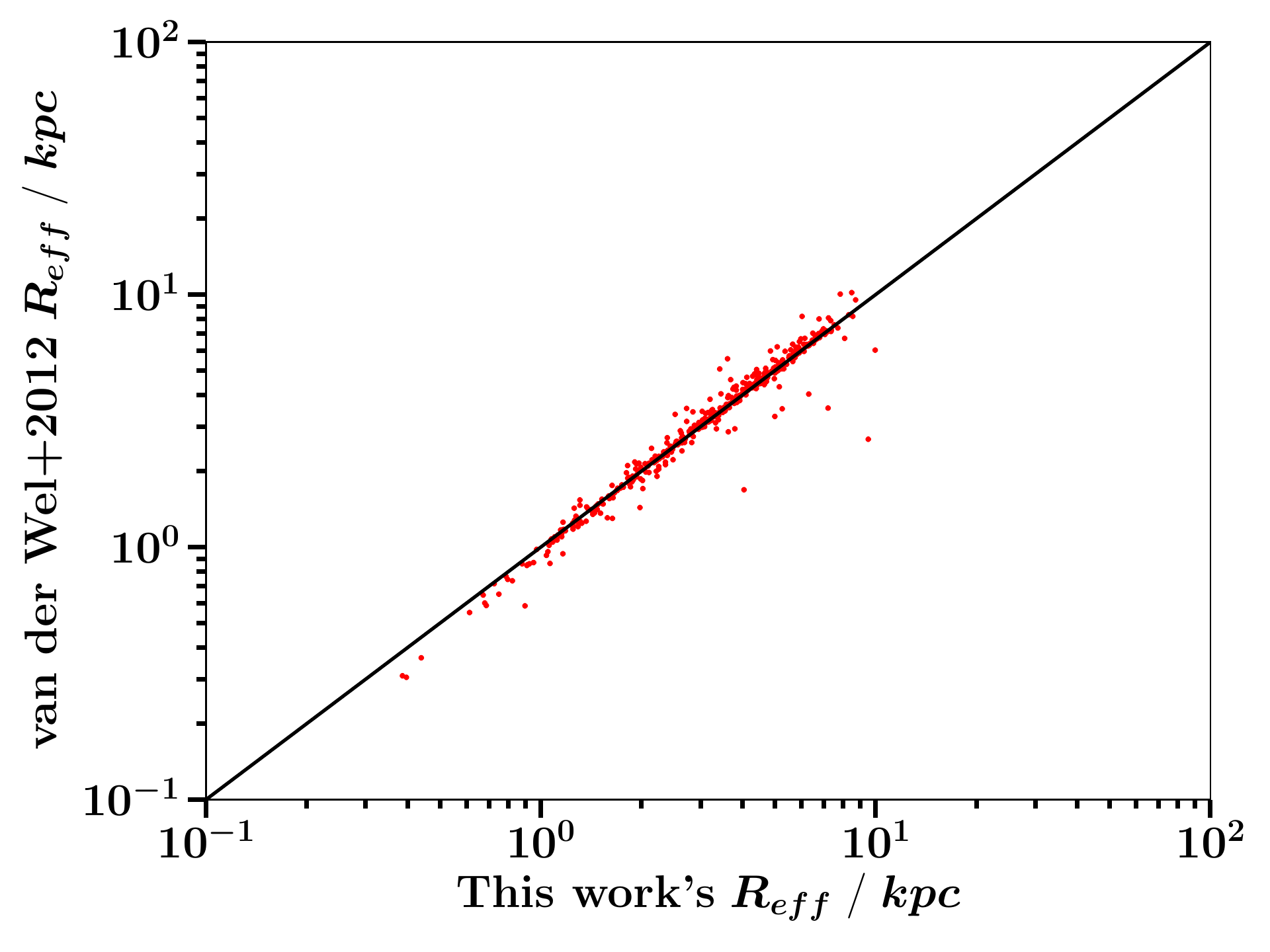}
    \caption{Level of agreement with results from \protect\cite{VanderWel2012} for the half-light radii measurements of the same set of galaxies with  \mbox{$0.86<z<1.34$}, \mbox{F160W magnitude~$<25$}, $R_{eff}<50$~kpc and stellar masses within the completeness limits of our study (log$(M_{*}/M_{\odot})>9.96$ and log$(M_{*}/M_{\odot})>9.60$ for quiescent and star-forming galaxies, respectively) from the CANDELS-COSMOS F160W mosaic. Solid line indicates the position of one-to-one agreement. The mean offset between the two measurements is $0.28\%$.}
    \label{fig:size_size}
\end{figure}

\section{The effect of differing filters on measured sizes} \label{filter_difference}

\begin{figure}
	\includegraphics[width=\columnwidth]{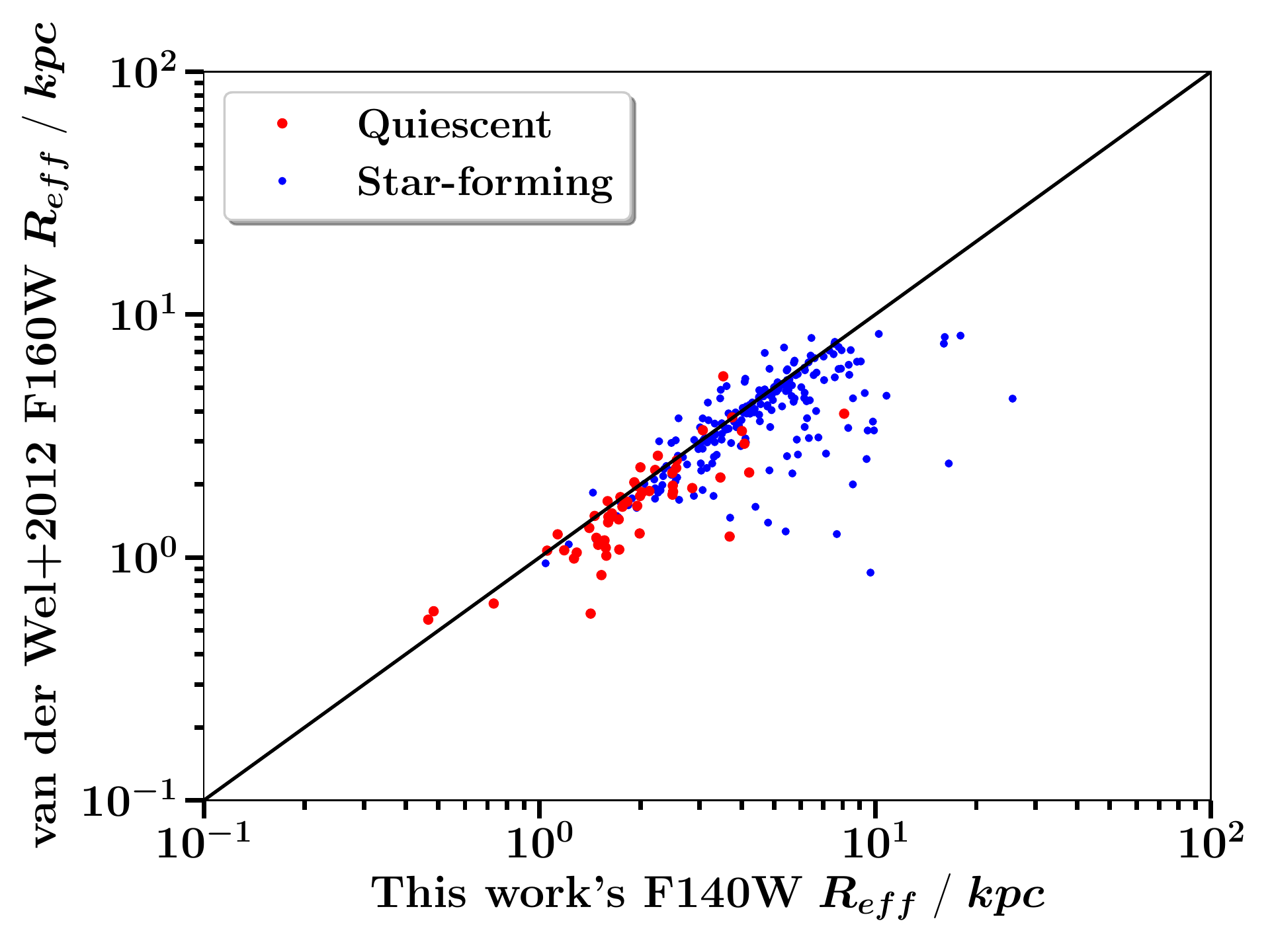}
    \caption{Level of agreement between the half-light radius measurements made by \protect\cite{VanderWel2012} in F160W versus those made for the same set of galaxies in F140W using our size determination method. Quiescent galaxies are shown as larger red points and star-forming galaxies are shown as smaller blue points. These galaxies have redshifts within the range \mbox{$0.86<z<1.34$}, \mbox{F160W and F140W magnitude~$<25$}, $R_{eff}<50$~kpc and stellar masses within the completeness limits of our study (log$(M_{*}/M_{\odot})>9.96$ and log$(M_{*}/M_{\odot})>9.60$ for quiescent and star-forming galaxies, respectively). The solid line indicates the position of one-to-one agreement. Galaxies are on average 12.87\% larger in F140W than in F160W. Quiescent and star-forming galaxies are 12.12\% and 13.04\% larger in F140W than in F160W respectively.}
    \label{fig:filter}
\end{figure}

As mentioned in the Introduction (Section~\ref{introduction}), colour gradients can vary for galaxies depending on which filter is used to observe them. This can lead to half-light radii measurements that differ for the same galaxy, depending on the filter used. Since the half-light radii measurements for our field sample were made using images in a different filter (F160W) to the filter used for the images of the cluster galaxies (F140W), we performed a check on how this affects conclusions drawn in this paper.

In Figure~\ref{fig:filter}, we plot our half-light radii measurements made from the 3D-HST COSMOS F140W mosaic against those made by \cite{VanderWel2012} for the same set of galaxies with  \mbox{$0.86<z<1.34$}, \mbox{F160W and F140W magnitude~$<25$}, $R_{eff} < 50$ kpc and stellar masses within the mass completeness limits of our study (log$(M_{*}/M_{\odot})>9.96$ and log$(M_{*}/M_{\odot})>9.60$ for quiescent and star-forming galaxies, respectively) from the CANDELS-COSMOS F160W mosaic. After taking into account the small systematic difference between the two size determination methods (see Appendix~\ref{size_agreement}), we find that galaxies are on average 12.87\% larger in F140W than in F160W. Quiescent and star-forming galaxies are 12.12\% and 13.04\% larger in F140W than in F160W respectively. Therefore, the sizes for the field sample used in this study would be on average 12.87\% larger if they were measured in F140W as opposed to F160W. This would therefore lead to a larger negative offset in the average size of cluster galaxies with respect to field galaxies, further strengthening the conclusions made in this paper.


\bsp	
\label{lastpage}
\end{document}